\documentclass[lettersize]{jfm}
\usepackage{amsmath,amsfonts,amssymb,amsbsy,amscd,amsgen}
\usepackage{subfigure}
\usepackage{float}
\usepackage[longnamesfirst]{natbib}
\usepackage{color,graphicx}

\newcommand{\refeqn}[1]{(\ref{#1})}

\newcommand{\refFigs}[1]{Figs.\ \ref{#1}}
\newcommand{\reffig}[1]{fig.\ \ref{#1}}
\newcommand{\refFig}[1]{Fig.\ \ref{#1}}
\newcommand{\refsec}[1]{\S\ \ref{#1}}
\newcommand{\refSec}[1]{Sec.\ \ref{#1}}
\newcommand{\bu}{\ensuremath{{\bf u}}}
\newcommand{\be}{\ensuremath{{\bf e}}}
\newcommand{\butot}{\ensuremath{{\bf u}_{\text{tot}}}}
\newcommand{\grad}{\ensuremath{{\bf \nabla}}}
\newcommand{\lapl}{\ensuremath{\nabla^2}}
\newcommand{\bx}{\ensuremath{{\bf x}}}

\newcommand{\Rey}{\ensuremath{\mathrm{Re}}}

\newcommand{\dd}[2]{\ensuremath{\frac{\partial #1}{\partial #2}}}

\newcommand{\tx}{\ensuremath{\tau_x}}

\newcommand{\txz}{\ensuremath{\tau_{xz}}}
\newcommand{\sx}{\ensuremath{\sigma_{x}}}
\newcommand{\sy}{\ensuremath{\sigma_{y}}}
\newcommand{\sz}{\ensuremath{\sigma_{z}}}
\newcommand{\sxy}{\ensuremath{\sigma_{xy}}}

\newcommand{\sxyz}{\ensuremath{\sigma_{xyz}}}

\newcommand{\xymeanu}{\langle u \rangle_{xy}}

\newcommand{\less}{\leq}

\newcommand{\greater}{\geq}

\graphicspath{{./figs-clr-png/}}

\title[Homoclinic snaking in plane Couette flow]{Homoclinic snaking in plane Couette flow:
bending, skewing, and finite-size effects
}

\author[John F. Gibson and Tobias M. Schneider]{
J.\ns F.\ns G\ls I\ls B\ls S\ls O\ls N,$^1$
\and 
T.\ns M.\ns S\ls C\ls H\ls N\ls E\ls I\ls D\ls E\ls R$^2$
}
\affiliation{
$^1$Department of Mathematics and Statistics, University of New Hampshire,
Durham, NH 03824 USA\\
[\affilskip]
$^2$Emergent Complexity in Physical Systems Laboratory (ECPS), 
\'Ecole Polytechnique F\'ed\'erale de Lausanne, CH-1015 Lausanne, Switzerland}

\date{\today}

\begin{document}

\maketitle


\begin{abstract}
Invariant solutions of shear flows have recently been extended from spatially periodic 
solutions in minimal flow units to spatially localized solutions on extended domains. 
One set of spanwise-localized solutions of plane Couette flow exhibits homoclinic 
snaking, a process by which steady-state solutions grow additional structure 
smoothly at their fronts when continued parametrically. Homoclinic snaking 
is well understood mathematically in the context of the one-dimensional Swift-Hohenberg 
equation. Consequently, the snaking solutions of plane Couette flow form a promising connection 
largely phenomenological study of laminar-turbulent patterns in viscous shear flows 
and the mathematically well-developed field of pattern-formation theory. In this paper we present 
a numerical study of the snaking solutions, generalizing beyond the fixed streamwise wavelength 
of previous studies. We find a number of new solution features, including bending, skewing, and 
finite-size effects. We show that the finite-size effects result from the shift-reflect symmetry 
of the traveling wave and establish the parameter regions over which snaking occurs.
A new winding solution of plane Couette flow is derived from a strongly skewed localized equilibrium. 

\end{abstract}


\section{Introduction}

Invariant solutions of the Navier-Stokes equations are known to play an important role 
in the dynamics of turbulence at low Reynolds numbers \citep{KawaharaARFM12}. These solutions, 
in the form of equilibria, traveling waves, and periodic orbits, have been computed precisely 
for canonical shear flows such as pipe flow \citep{FaisstPRL03, WedinJFM04, DuguetPF08},
plane Couette flow \citep{NagataJFM90,KawaharaJFM01,ViswanathJFM07,GibsonJFM09} 
and plane Poiseuille flow \citep{WaleffeJFM01,GibsonJFM14}. The development of the invariant-solutions 
approach to turbulence has largely occurred in the simplified context of small, periodic domains, 
or `minimal flow units' \citep{JimenezJFM91}. More recently, invariant solutions with localized 
support have been computed for flows on spatially extended domains. 
These include 
spanwise-localized equilibria and traveling waves \citep{SchneiderJFM10,SchneiderPRL10,DeguchiJFM13,GibsonJFM14} 
in plane Couette flow, and 
spanwise-localized traveling waves \citep{GibsonJFM14} and a periodic orbit 
\citep{ZammertFDR14} of plane Poiseuille flow.
\cite{AvilaPRL13} computed a streamwise-localized periodic orbit of pipe flow, 
\cite{MellibovskyJFM15} a streamwise-localized periodic orbit of plane Poiseuille flow, 
\cite{BrandJFM14} a doubly-localized equilibrium solution of plane Couette flow, and
\cite{ZammertJFM14} a doubly-localized periodic orbit of plane Poiseuille flow.
The existence and structure of these spatially localized solutions suggests that they are 
relevant to large-scale patterns of laminar-turbulent intermittency, such as turbulent 
stripes, spots, and puffs. For example, the periodic orbit of \cite{AvilaPRL13} shares the 
spatial structure and complexity of turbulent puffs in pipe flow, and its bifurcation 
sequence provides a compelling explanation of the development of transient turbulence in pipes.
The doubly-localized equilibrium of \cite{BrandJFM14} has the characteristic shape and structure 
of turbulent spots in low-Reynolds plane Couette flow, and for a range of Reynolds numbers sits 
on the boundary between laminar flow and turbulence. 
Analysis of these localized solutions has so far focused on their bifurcations from 
spatially periodic solutions \citep{ChantryPRL14,MellibovskyJFM15} and linear analysis 
of their decaying tails \citep{BrandJFM14,GibsonJFM14}. 

The spanwise-localized invariant solutions of plane Couette flow of \cite{SchneiderJFM10}, 
are notable for being the first localized solutions discovered, for their relation to the 
widely-studied equilibrium solution of \cite{NagataJFM90,CleverJFM97,WaleffePRL98} 
(hereafter NBCW), and for exhibiting the particularly interesting feature of 
homoclinic snaking. 
Homoclinic snaking is a process by which the localized solutions 
grow additional structure at their fronts in a sequence of saddle-node bifurcations when 
continued parametrically (\cite{KnoblochARCMP15,SchneiderPRL10}; see also 
\refsec{sec:snaking}). 
Homoclinic snaking occurs a number of pattern-forming systems with 
localized solutions, including binary fluid convection \citep{BatistePRL05} and 
magneto-convection \citep{BatisteJFM06}, and it is well-understood mathematically for 
the one-dimensional Swift-Hohenberg equation \citep{BurkeChaos07,BeckSJMA09}.  
\cite{KnoblochARCMP15} provides a comprehensive review of localization and homoclinic 
snaking in dissipative systems. Though no explicit connection between the Swift-Hohenberg
and the Navier-Stokes equations is known, the striking similarity of the localized plane 
Couette solutions and the localized solutions of Swift-Hohenberg with cubic-quintic 
nonlinearity suggests there might be a mathematical connection between the two systems. 
These similarities include the structure of localization, the snaking behaviour, 
the even/odd symmetry of the snaking solutions, and the existence of asymmetric rung 
solutions \citep{SchneiderPRL10}. 
One might envision, for example, that a reduced-order model of the localized solutions
\citep{HallJFM10,HallJFM12,BeaumePRE15} might relate the spanwise variation of 
their mean streamwise flow to the cubic-quintic Swift-Hohenberg equation.
Such a relation would link the mathematically well-developed field of pattern-formation 
theory to the localized solutions of shear flows cited above, or to recent 
numerical studies of laminar-turbulent pattern formation in extended 
shear flows \citep{BarkleyPRL05,DuguetJFM10,TuckermanPF14}.

In support of developing such connections between pattern-formation theory and
shear flows, we present in this paper a more detailed analysis of the snaking solutions 
of \cite{SchneiderJFM10,SchneiderPRL10}. In particular, we examine the effects of varying 
the streamwise wavelength $L_x$ of the solutions compared to the fixed $L_x = 4\pi$ of 
\cite{SchneiderPRL10}.  We find that homoclinic snaking is robust in $L_x$ and that the 
snaking region moves upwards in Reynolds number with decreasing $L_x$. The ranges of 
streamwise wavelength and Reynolds number in which snaking solutions exist is found 
to be $1.7\pi \less L_x \less 4.2\pi$ and  $165 \less \Rey \less 2700$. 
Additionally, we find several interesting solution properties that are suppressed 
at the parameters studied in \cite{SchneiderPRL10}. As 
$L_x$ as decreases below $4\pi$ and $\Rey$ increases above $165$, the localized 
solutions deform appreciably compared to their strictly periodic counterparts, 
the localized equilibria exhibiting a linear skewing and the traveling waves a quadratic 
bending. We show that skewing and bending are related to the respective symmetries of 
the equilibrium and traveling wave solutions, and that bending induces finite-size 
effects in the traveling waves that scale as the inverse of their spanwise width. 
In contrast, skewing induces no such finite-size effects on the equilibrium solution. 
We show that the skewed solutions lead to a new periodically winding form of the 
NBCW equilibrium solution of plane Couette flow. 

The structure of this paper is as follows. \refSec{sec:methods} outlines the 
problem formulation and numerical methods. \refSec{sec:fixedLx} describes the 
features of the localized solutions at fixed streamwise wavelength $L_x$, 
including homoclinic snaking, bending, skewing, and finite-size effects.
\refSec{sec:varyingLx} discusses the effects of varying streamwise wavelength, 
including the regions of wavelength and Reynolds number over which snaking occurs,
the breakdown of snaking outside these regions, and the stability of the
solutions. \refSec{sec:periodicpattern} discusses the periodic pattern in
the interior of the localized solutions and its relation to the NBCW solution.
The new winding solution is also presented in \refsec{sec:periodicpattern}.

\section{Problem formulation, methodology, and conventions}
\label{sec:methods}

Plane Couette flow consists of an incompressible Newtonian fluid between two 
infinite parallel plates moving at constant relative velocity. The Reynolds 
number is given by $\Rey = U h/\nu$ where $U$ is half the relative wall speed, 
$h$ is half the distance between the walls, and $\nu$ is the kinematic viscosity. 
The $\bx = (x,y,z)$ coordinates are aligned with the streamwise, wall-normal, and 
spanwise directions, where streamwise is defined as the direction of relative wall 
motion. After nondimensionalization the walls at $y=\pm 1$ move at speeds $\pm 1$
in the $x$ direction, and the laminar velocity field is given by $y \be_x$. We 
decompose the total fluid velocity into a sum of the laminar flow and the deviation 
from laminar: $\butot = y \be_x + \bu$. Hereafter we refer to the deviation field 
$\bu(\bx,t) = [u,v,w](x,y,z,t)$ as ``velocity.'' In these terms the laminar solution 
is specified by $\bu = 0$, $p=0$ and the Navier-Stokes equations take the form
\begin{align}
\dd{\bu}{t} + y \dd{\bu}{x} + v \be_x + \bu \cdot \grad \bu  &= - \grad p + \frac{1}{\Rey} \lapl \bu, \quad \grad \cdot \bu = 0.
\label{eqn:NS}
\end{align}
The computational domain $\Omega = [-L_x/2, \: L_x/2] \times [-1, 1] \times [-L_z/2, \: L_z/2]$ 
has periodic boundary conditions in $x$ and $z$ and no-slip conditions at the walls.
For spanwise-localized solutions, $L_z$ is typically large, so that $\Omega$ approximates 
a spanwise-infinite domain. We use $\hat{L}_z$ to denote the spanwise wavelength of 
nearly periodic, small-wavelength patterns within the spanwise-localized solutions; typically
$\hat{L}_z \ll L_z$. 
In the present work we impose zero mean pressure gradient in all computations, leaving the 
mean (bulk) flow to vary dynamically. As described in \cite{GibsonJFM08,GibsonJFM09}, direct 
numerical simulations are performed with Fourier-Chebyshev spatial discretization and semi-implicit 
time-stepping, traveling-wave and equilibrium solutions of \refeqn{eqn:NS} are computed with
a Newton-Krylov-hookstep algorithm, and all software and solution data is available for download 
at {\tt www.channelflow.org}.

The equilibrium and traveling-wave solutions discussed here are all steady states 
(in a fixed or traveling frame of reference, respectively), so the energy dissipation rate 
balances the power input from wall shear instantaneously:
\begin{align}
D = I = \frac{1}{2L_x} \int_{-L_x/2}^{L_x/2} \int_{-L_z/2}^{L_z/2} \left. \dd{u}{y} \right|_{y=-1} + \left. \dd{u}{y} \right|_{y=1} dx \, dz.
\label{eqn:D}
\end{align}
Note that $D$ is defined in terms of the deviation velocity $\bu$ and not the total velocity 
$\butot$, so that $D$ measures the excess energy dissipation of spanwise-localized solutions 
over the laminar flow, which has $D=0$. Since the internal structure of a spanwise-localized
solutions stays roughly constant as non-laminar structure grows at its fronts, $D$ serves as 
a good measure of the width of a solution. The lack of $L_z$ normalization makes the $D$ of a 
spanwise-localized solution insensitive to the choice of spanwise length for the 
computational domain in which it is embedded.

For discussing the symmetries of the flow we follow the conventions of \cite{GibsonJFM14},
here adding the action of symmetries on the pressure field. Let
\begin{align}
\sx &: [u,v,w,p](x,y,z) \rightarrow [-u,v,w,p](-x,y,z), \nonumber\\
\sy &: [u,v,w,p](x,y,z) \rightarrow [u,-v,w,p](x,-y,z), \nonumber\\
\sz &: [u,v,w,p](x,y,z) \rightarrow [u,v,-w,p](x,y,-z), \nonumber\\
\tau(\Delta x, \Delta z) &: [u, v, w,p](x,y,z) \rightarrow [u, v, w,p](x+\Delta x,\, y,\, z+\Delta z), \label{eq:sigmadef}
\end{align}
and let concatenation of subscripts indicate products, e.g.\ $\sxy = \sx \sy$. 
For $(\ell_x, \ell_z)$-periodic fields we define two half-wavelength translation operators
$\tau_x = \tau(\ell_x/2, 0)$ and $\tau_z = \tau(0, \ell_z/2)$. The standard
group-theoretic angle-bracket notation indicates the group formed by a set of 
generators; for example $\langle \sxy, \tx \sz  \rangle = \{e,\, \sxy,\, \tx \sz,\, \tx \sxyz \}$,
where $e$ is the identity \citep{DummitFoote04}. 



\section{Solution properties at fixed streamwise wavelength}
\label{sec:fixedLx}

\subsection{Snaking}
\label{sec:snaking}

\begin{figure}
\begin{center}
\footnotesize{(a)} \hspace{-2mm} \includegraphics[width=2.4in]{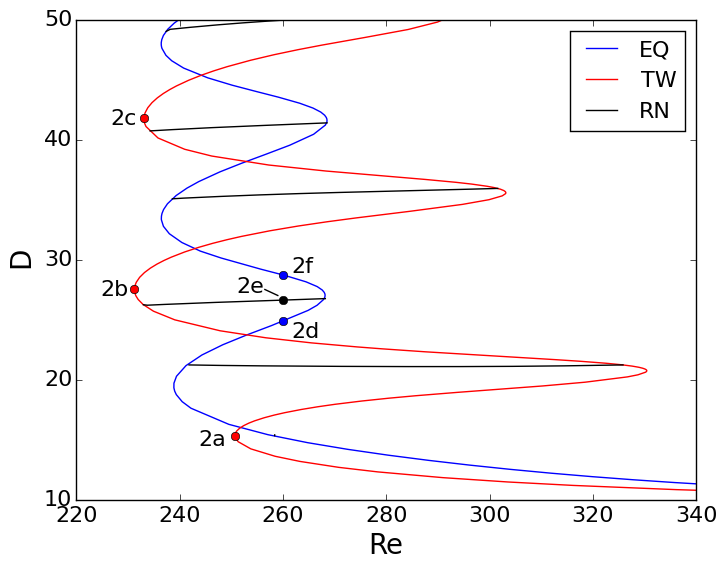}
\footnotesize{(b)} \hspace{-2mm} \includegraphics[width=2.4in]{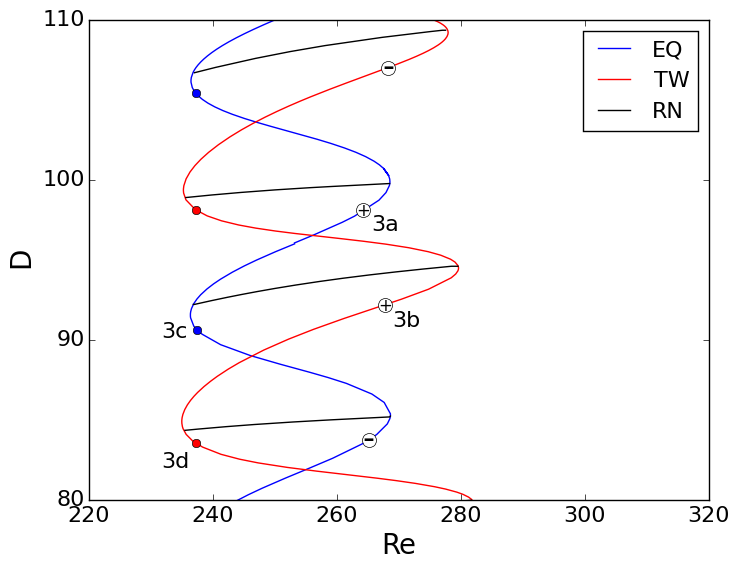}
\caption{{\bf Homoclinic snaking of localized solutions at $L_x=3\pi$} (details). 
(a) Snaking curves at low $D$ (small spanwise width) for localized equilibrium (EQ), 
traveling-wave (TW), and rung (RN) solutions. Labels indicate the solutions
shown as velocity fields in \reffig{fig:velocity}.
(b) Snaking curves at high $D$ (large spanwise width). Filled circles
indicate the points of zero skewing (for equilibria) and zero bending 
(for traveling waves) along the snaking curves; signed open circles mark the 
positions and signs of the maxima in magnitude of skewing and bending 
(see \refsec{sec:bendskew}). Pressure fields for the labeled points are 
shown in \reffig{fig:bendskew}. Both subplots are details of the $L_x=3\pi$ 
snaking curve shown in \reffig{fig:snaking_threepi}(d). 
\label{fig:snaking_details}
}
\end{center}
\end{figure}

The primary notable feature of the localized solutions is their {\em homoclinic snaking}. 
Under continuation in Reynolds number at fixed streamwise wavenumber, the localized 
equilibrium and traveling-wave solutions follow curves that snake upwards in the 
$\Rey,D$ plane, as shown in detail in \reffig{fig:snaking_details} and over a larger range
of $D$ in \reffig{fig:snaking_threepi}(d). Velocity fields corresponding to the labeled 
points in \reffig{fig:snaking_details}(a) are shown in \reffig{fig:velocity}. 
\refFigs{fig:velocity}(a,b,c) show that the traveling-wave solution grows additional
structure at the solution fronts as it moves upwards in $D$ along the snaking curve, 
while the interior structure remains nearly constant. The structure of the fronts 
is the same at alternating saddle-node points (a,c), 
while the saddle-node point (b) between them has front structure of opposite streamwise sign. 
Note that due to the $\sxy$ symmetry of plane Couette flow, every traveling-wave solution 
$\bu$ with wave speed $c_x$ has a symmetric partner $\sxy \bu$ with wave speed $-c_x$. 
The  $\sxy \bu$ symmetric partner of \reffig{fig:velocity}(b) has fronts with the same structure
and streamwise sign as \reffig{fig:velocity}(a,c).

\refFig{fig:snaking_details}(a) also shows ``rung'' solutions that bifurcate from the 
equilibrium solution in a pitchfork bifurcation near the saddle-node bifurcation 
points of the equilibrium and connect to the traveling wave near their saddle-node
points (or vice versa). The existence of the rung solutions can 
be understood from a physical viewpoint as a combination of two solutions near the 
saddle-node bifurcation point, with different widths $D$ but the same Reynolds number.
For example, the equilibria marked 2d and 2f in \reffig{fig:snaking_details}(a) and 
depicted as streamwise velocity fields in \reffig{fig:velocity}(d,f) are indistinguishable 
within the interior $-5 < z < 5$. But their differing values of $D$ indicate different 
spanwise lengths. The contour lines of the fronts of 2(f) extend towards $|z| \approx 10$, 
whereas those of 2(d) reach just $|z| \approx 8$. The rung solution shown as 
\reffig{fig:velocity}(e) and marked 2e in \reffig{fig:snaking_details}(a) can then be 
understood as splicing together the left half of \reffig{fig:velocity}(d) and the 
right half of \reffig{fig:velocity}(f). This splicing can be done at arbitrary $\Rey$ 
in the interior of the saddle-node bifurcation, i.e. along the black lines 
of the rung branches shown \reffig{fig:snaking_details}(a). The splicing construction
is necessarily imperfect, since the rung solutions have no symmetries and hence travel 
in both $x$ and $z$, compared to the equilibrium, which is fixed. However it is close 
enough that such spliced velocity fields converge quickly to the rung solutions under 
Newton-Krylov-hookstep search. The rung solutions in this paper were computed
by splicing and refinement, followed by continuation in Reynolds number. 

\newcommand{\lbl}[1]{\footnotesize{(#1)} \hspace{-2mm}} 
\newcommand{\phn}[1]{\hspace{#1mm}} 

\begin{figure}
\begin{center}
\phn{2.5} \includegraphics[width=2.45in]{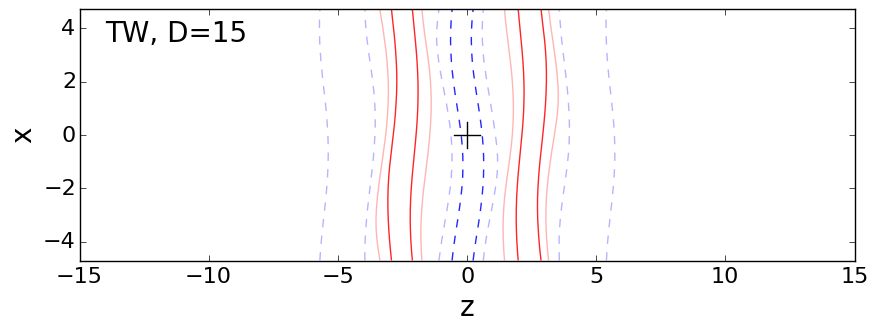}  \phn{3.2} \includegraphics[width=2.45in]{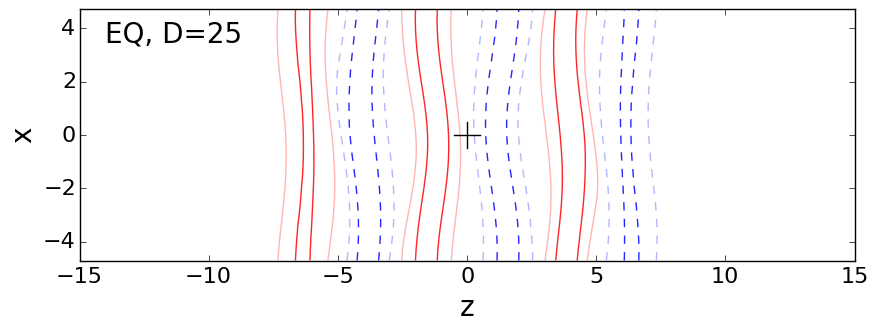} \\
\lbl{a}   \includegraphics[width=2.45in]{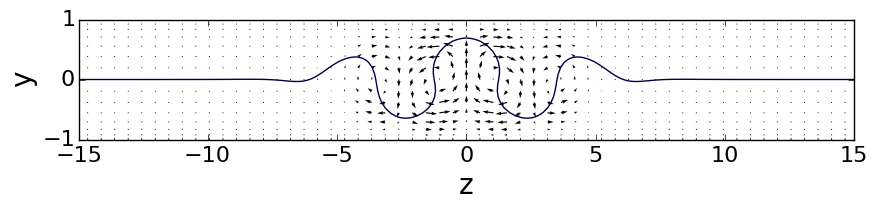} \lbl{d}   \includegraphics[width=2.45in]{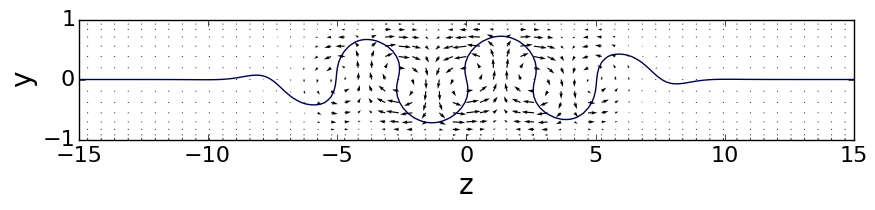} \\ 
\phn{2.5} \includegraphics[width=2.45in]{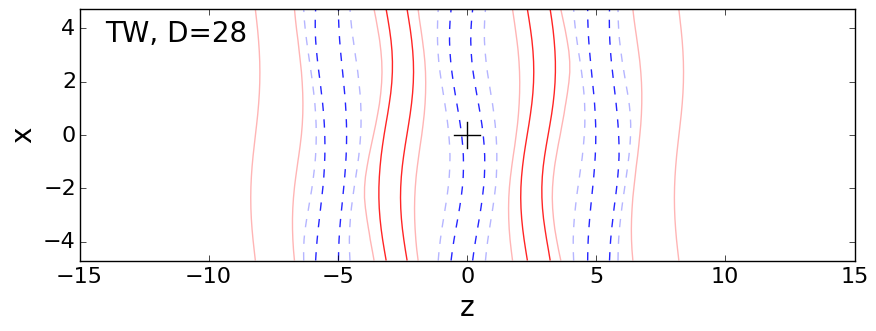}  \phn{3.2} \includegraphics[width=2.45in]{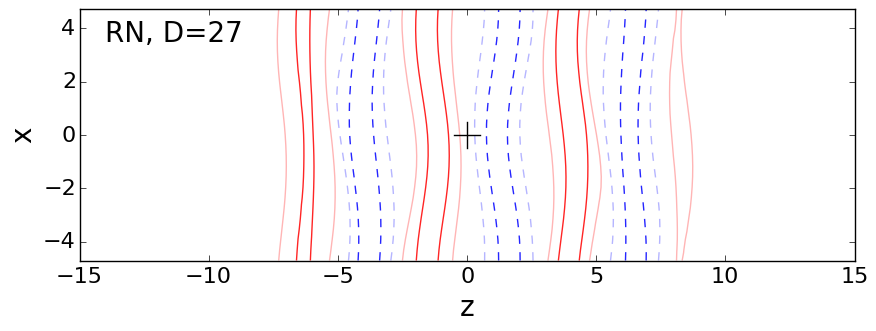} \\
\lbl{b} \includegraphics[width=2.45in]{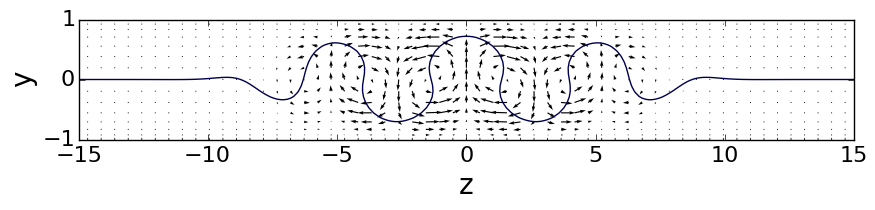}   \lbl{e}   \includegraphics[width=2.45in]{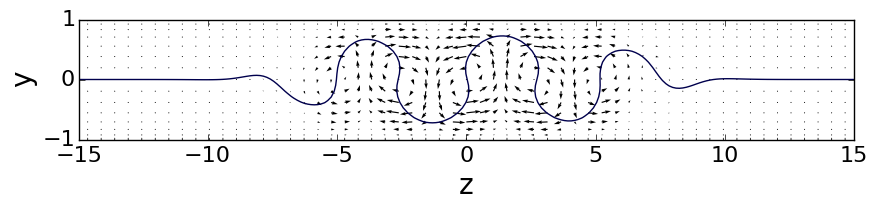} \\
\phn{2.5} \includegraphics[width=2.45in]{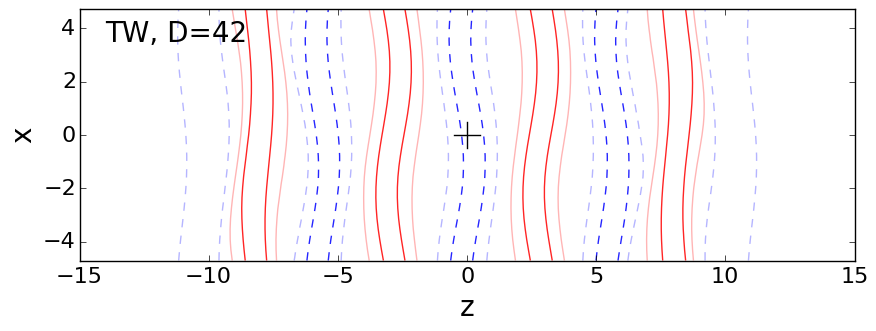}  \phn{3.2} \includegraphics[width=2.45in]{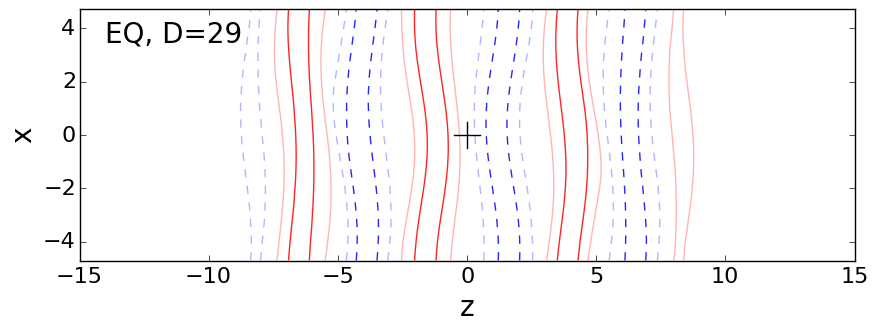} \\
\lbl{c}   \includegraphics[width=2.45in]{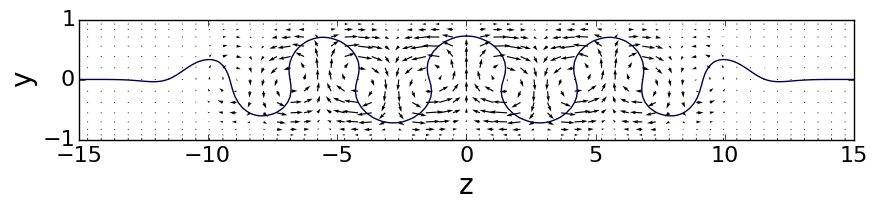} \lbl{f}   \includegraphics[width=2.45in]{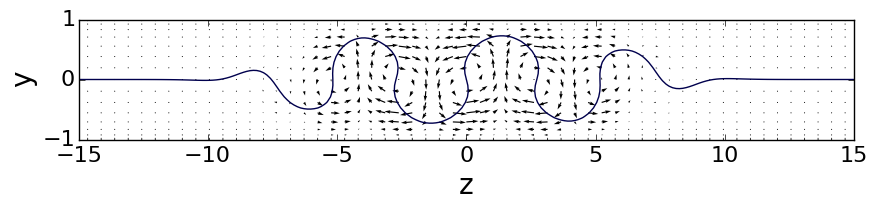} 
\end{center}
\caption{(color online) {\bf Velocity fields of localized solutions} 
illustrated by contours of streamwise velocity in the $y=0$ midplane, $u(x,0,z)$ and 
arrow plots of the streamwise-averaged cross-stream velocity, $[\bar{v},\bar{w}](y,z)$.
Eight contour levels are evenly spaced between $\pm0.9$, with negative $u$ 
in dashed blue lines and positive in solid red.
Contour lines for $\bar{u}=0$ are superimposed on the arrow plots. (a,b,c) show the traveling-wave 
(TW) solution at the three successive lower saddle-node bifurcation points marked on the snaking 
curve in \reffig{fig:snaking_details}(a). (e,f) show the equilibrium (EQ)
solution at $\Rey=260$ above and below an upper saddle-node bifurcation, and (d) shows 
the rung solution at $\Rey=260$, at points labeled in \reffig{fig:snaking_details}(a).
The solution is shown for $L_x=3\pi$ and on a subset of the $L_z=16\pi$ 
computational domain. 
}
\label{fig:velocity} 
\end{figure}

\subsection{Symmetries of localized solutions}
\label{sec:symmetry}

The differences between traveling waves, equilibria, and rungs are intimately related 
to the different symmetries of those solutions, which can be understood in terms of 
symmetry-breaking bifurcations of the more symmetric, spatially periodic NBCW solution. 
This is discussed in detail in \cite{GibsonJFM14}; here we present a brief summary. 
With proper placement of the $z$ origin, the traveling waves have a $\tx \sz$ 
``shift-reflect'' symmetry. That is, a traveling-wave solution satisfies 
$\bu = \tx \sz \bu$ or 
\begin{align}
[u,v,w,p](x,y,z) = [u,v,-w,p](x+\ell_x/2, y, -z).
\label{eqn:twsymm}
\end{align}
Solutions with this symmetry can travel in $x$ but not $z$, since the inversion
in $z$ about the origin locks the $z$ phase of the solution, but no 
such restriction exists for $x$. For similar reasons, the traveling waves can have
nonzero mean streamwise velocity, but their mean spanwise velocity must be zero.
The $\tx \sz$ symmetry of the localized traveling waves arises from a subharmonic-in-$z$ 
bifurcation of the $(\ell_x, \ell_z)$-periodic NBCW solution, which in the spatial phase of 
\cite{WaleffePF03}, has symmetries $\langle \tx \sz, \txz \sxy \rangle$.  
The subharmonic-in-$z$ bifurcation necessarily breaks the $\txz \sxy$ 
symmetry, since this symmetry implies $\ell_z$ periodicity, as follows. 
If $\txz \sxy \bu = \bu$, then $(\txz \sxy)^2 \bu = \bu$. But a brief calculation
shows that $(\txz \sxy)^2 = \tau(0,\ell_z)$. Thus the bifurcated solution loses
the  $\txz \sxy$ symmetry of NBCW and retains only $\tx \sz$.

The localized equilibrium solution has $\sxyz$ inversion symmetry, satisfying
\begin{align}
[u,v,w,p](x,y,z) = [-u,-v,-w,p](-x, -y, -z).
\label{eqn:eqsymm}
\end{align}
As a result of the inversion of all velocity components about the origin, the 
spanwise-localized solutions with this symmetry is prevented from traveling in $x$ 
or $z$, and the spatial average of all velocity components is zero. The $\sxyz$ 
symmetry of the localized equilibrium arises from a similar bifurcation of a 
phase-shifted NBCW solution. Shifting 
the NBCW solution by a quarter-wavelength in $z$, $\bu \rightarrow \tau_z^{1/2} \bu$, 
changes each of its symmetries $s$ to the conjugate symmetry $\tau_z^{-1/2} s \tau_z^{1/2}$
\citep{GibsonJFM09}. 
A brief calculation shows that the conjugated symmetry group of the phase-shifted NBCW 
solution is $\langle \txz \sxy, \sxyz \rangle$. The $\txz \sxy$ symmetry implies 
$\ell_z$-periodicity, as before, so the subharmonic-in-$z$ bifurcation breaks the 
$\txz \sxy$ symmetry but retains $\sxyz$. 

The symmetries of the traveling-wave and equilibrium solutions and the lack of symmetry in rung 
solutions are evident in the velocity-field plots shown in \reffig{fig:velocity}. The $z$-mirror, 
$x$-shift $\tx \sz$ traveling-wave symmetry \refeqn{eqn:twsymm} is particularly apparent in the fronts of the 
midplane $u$ contour plots of \reffig{fig:velocity}(a,b,c), and an even $z$-mirror symmetry 
is apparent in the corresponding $x$-averaged cross-stream $[\bar{v},\bar{w}](y,z)$ plots. 
It is also evident from these plots why the traveling-wave solution travels in $x$. 
In each of \reffig{fig:velocity}(a,b,c), both the $u(x,0,z)$ plots and the 
$[\bar{v},\bar{w}](y,z)$ plots show a clear imbalance between the positive/negative 
streamwise streaks. In comparison, for the equilibrium solutions, the $\sxyz$ symmetry 
of the equilibrium matches each streamwise streak at negative $z$ with an equal streak 
at positive $z$ of opposite sign. The rung solution \reffig{fig:velocity}(e), in contrast,
has no symmetry at all. The lack of symmetry in the rungs is due fundamentally to their 
symmetry-breaking bifurcations from the traveling-wave and equilibrium solutions. It can also 
be understood physically as a consequence of the formation of rungs via splicing as described 
in \refsec{sec:snaking}, which clearly breaks the $\sxyz$ symmetry of the equilibrium solution 
(or the $\tx \sz$ symmetry if constructed by splicing traveling waves). The complete lack of 
symmetry in rung solutions means they generally have nonzero wave speeds and nonzero net  
velocity in both the stream- and spanwise directions. 

\subsection{Bending, skewing, and finite-size effects}
\label{sec:bendskew}

\begin{figure}
\begin{center}
\lbl{a} \includegraphics[width=4.5in]{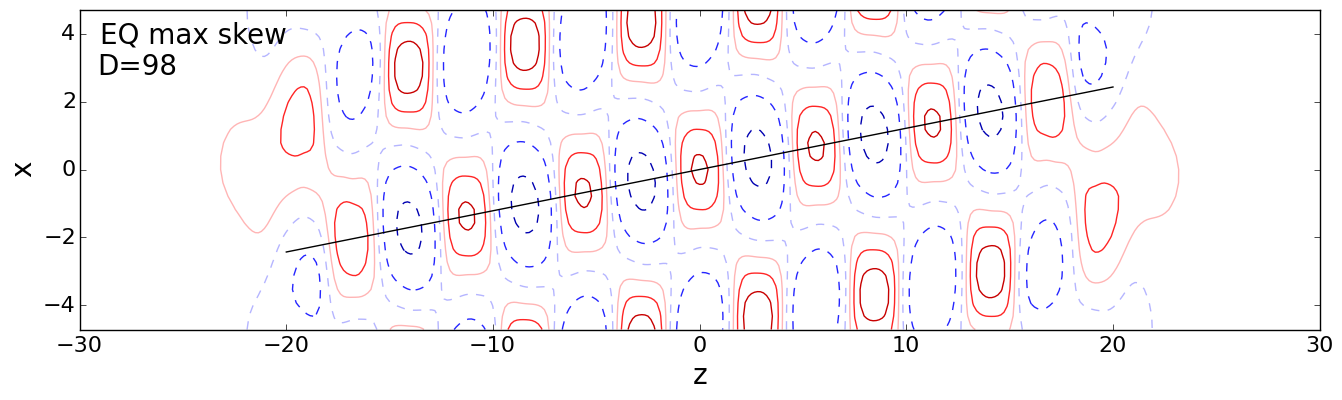}\\
\lbl{b} \includegraphics[width=4.5in]{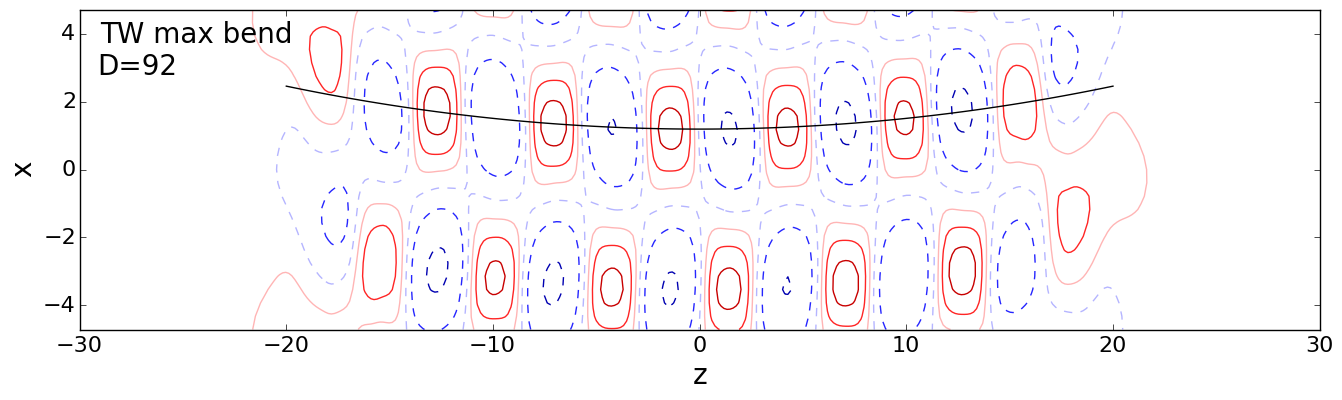}\\
\lbl{c} \includegraphics[width=4.5in]{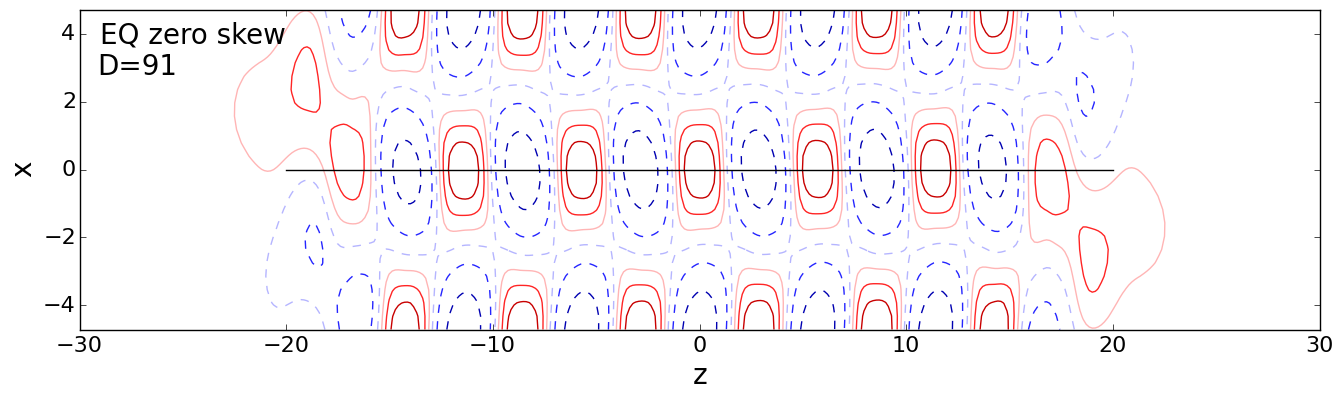}\\
\lbl{d} \includegraphics[width=4.5in]{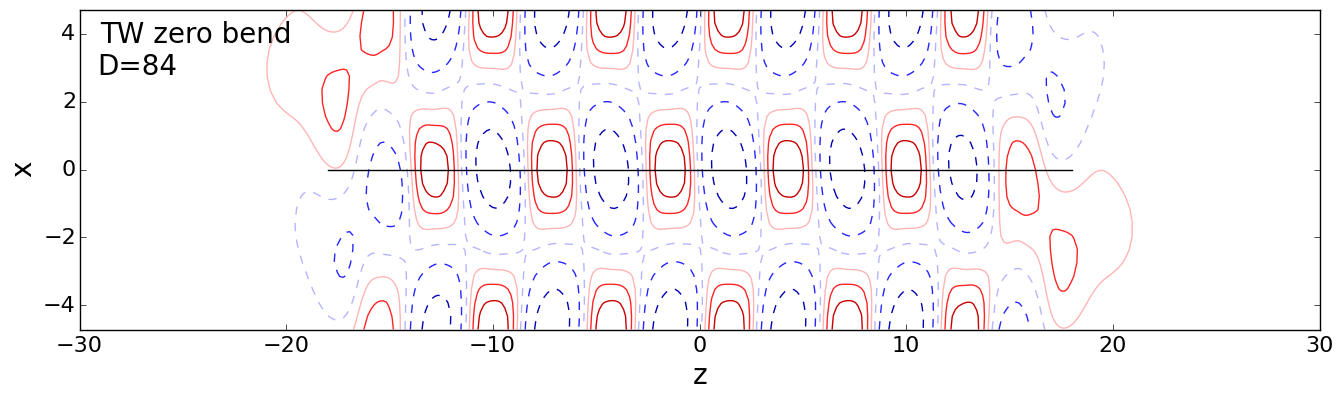}
\caption{{\bf Bending and skewing.} (color online)
Contour plots of pressure $p(x,0,z)$ in the $y=0$ midplane are shown for localized 
equilibrium (EQ) and traveling-wave (TW) solutions with maximum and zero skewing and 
bending, corresponding to points marked on \reffig{fig:snaking_details}(b). 
In (a,c), a line of constant slope passes through the pressure minima 
and maxima, showing uniformity of skew throughout the interior of the equilibrium 
solution, with (a) showing maximum skewing and (c) zero skewing. 
In (b,d), a line of constant curvature shows the uniformity of bending 
for the traveling wave, with (b) showing maximum bending and (d) zero bending.
Eight contour levels are evenly spaced between $p=\pm 0.025$, dashed 
blue for negative $p$ and solid red for positive. 
The solution is shown for $L_x=3\pi$ and on a subset of the $L_z=16\pi$ 
computational domain.
\label{fig:bendskew}
}
\end{center}
\end{figure}

\begin{figure}
\begin{center}
\footnotesize{(a)} \includegraphics[width=2.4in]{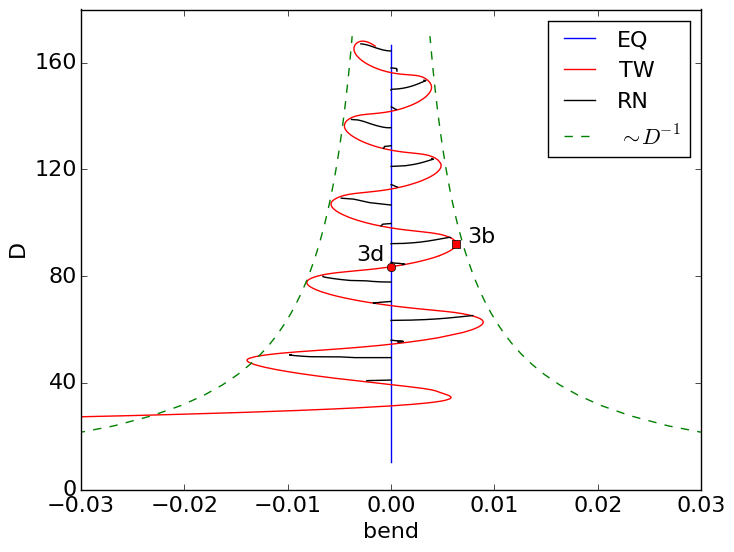}
\footnotesize{(b)} \includegraphics[width=2.4in]{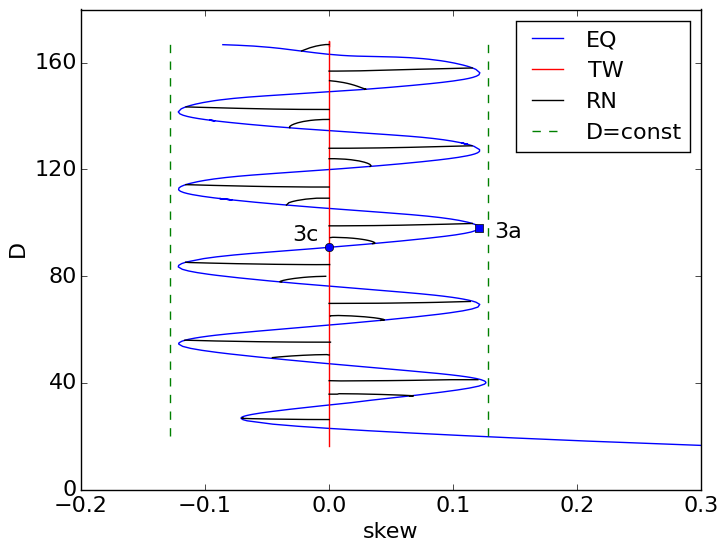}\\
\footnotesize{(c)} \includegraphics[width=2.4in]{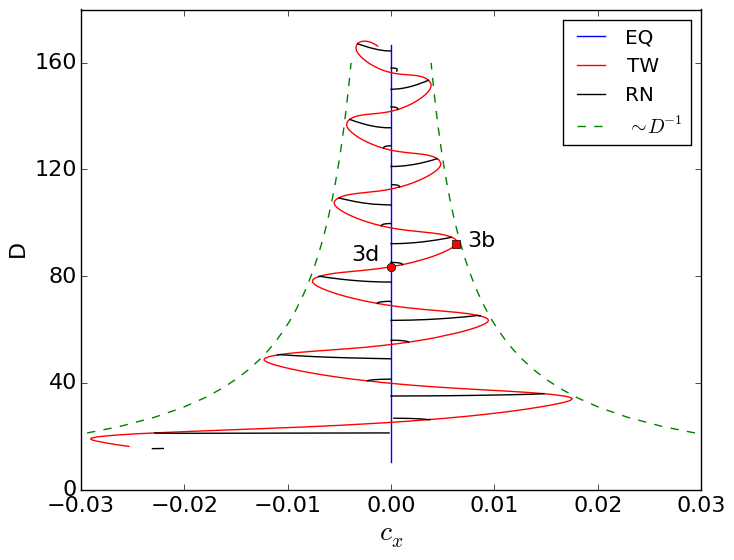} 
\footnotesize{(d)} \includegraphics[width=2.4in]{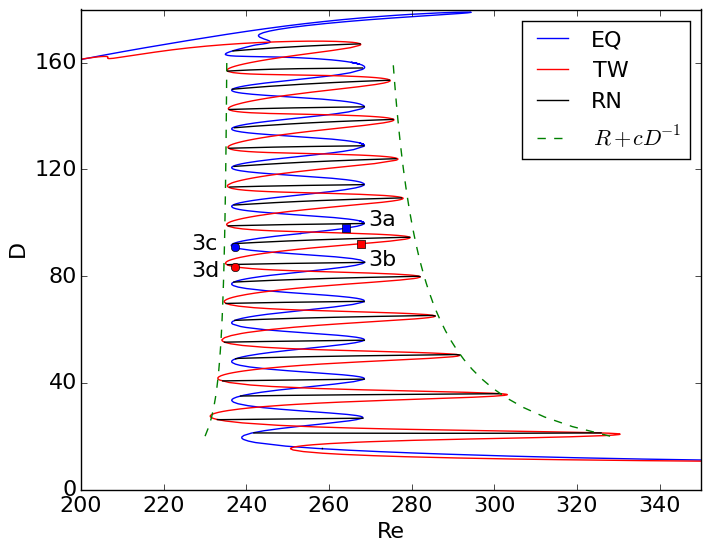}
\caption{{\bf (a) Bending, (b) skewing, and (c) wave speed in comparison to (d) snaking in Reynolds 
number} 
for the equilibrium (EQ), traveling-wave (TW), and rung (RN) solutions at $L_x=3\pi$ and a $L_z=24\pi$
computational domain. Dashed lines show the $~D^{-1}$ envelope of wave speed and 
bending for the traveling wave and the $D=\text{const}$ envelope of equilibrium skewing.
In (a) two independent dashed lines of form $R + cD^{-1}$ are shown.
The values of $R$ for the two lines were set as the lower and upper bounds of the equilibrium 
snaking curve ($R=236$ and $R=268$), and the values of $c$ chosen to fit the envelope 
of the traveling-wave snaking curve. 
Labeled points correspond to pressure fields shown in \reffig{fig:bendskew}. 
\label{fig:snaking_threepi}
}
\end{center}
\end{figure}

The equilibrium (EQ), traveling-wave (TW), and rung solutions shown as velocity fields in 
\reffig{fig:velocity} are at low $D$ and thus have small spanwise width. The three 
different types of solutions appear at first glance to consist of a few copies 
of the same spanwise-periodic structure placed side-by-side, with fronts on either 
side that taper to laminar flow. This description, however, is neither entirely accurate
nor complete. First of all, the interior structure of the three types of solutions 
must differ at least slightly because the solution types move at different wave speeds 
($c_x=c_z=0$ for equilibria, $c_x\neq 0, c_z=0$ for traveling waves, and 
$c_x \neq 0, c_z \neq 0$ for the rungs).
But further differences between the three solutions types become apparent at higher
$D$ and greater width. In this subsection we show that 
\begin{itemize}
\item the EQs {\em skew}, displaying a linear tilt in $x$ against $z$ (\reffig{fig:bendskew}a),
\item the TWs {\em bend}, displaying a quadratic curvature in $x$ against $z$ (\reffig{fig:bendskew}b),
\item the EQ snaking region has constant bounds in $\Rey$ (\reffig{fig:snaking_threepi}d),
\item the TW snaking region is wider but converges to the EQ's as $D^{-1}$ (\reffig{fig:snaking_threepi}d),
\item the TW's streamwise wavespeed decreases to zero as $D^{-1}$, (\reffig{fig:snaking_threepi}c),
\item the EQ's interior structure is periodic and winds in $x,z$ (\reffig{fig:bendskew}a), and
\item the TW's interior structure is nonperiodic and slowly modulated in $z$ (\reffig{fig:bendskew}b).
\end{itemize}
The common thread among these phenomena is the interplay between the fronts
and the interior structure. Much of the above can be understood by assuming 
that the fronts are the determining structures of the solutions, and viewing 
the other properties as a consequences of the fronts and their orientations, 
as determined by the solution symmetries. 

In this paragraph we present a brief sketch of the interplay between the fronts, 
symmetries, and solution properties. A fully detailed presentation follows 
in the remainder of the subsection. For the equilibrium, the 
odd symmetry and opposite orientation of the fronts about the origin produces a 
linear $x,z$ skew within the solution's interior. The uniform linear skew allows for 
periodic structure in the interior that winds linearly in $x,z$. The winding periodic 
structure oscillates with $D$, but is otherwise independent of the overall solution 
width. Consequently, many equilibrium solution properties are independent of the 
overall solution width. In contrast, for the traveling wave, the even $z$-mirror 
symmetry and similar orientation of the fronts produces quadratic $x,z$ bending 
in the interior. This curvature necessarily breaks the periodicity of the solution's 
interior structure and couples the interior structure and global properties to the 
solution width. The wave speed, bending, snaking region, and interior modulation of 
the traveling wave all vary according to the relative size of the fronts to the 
overall solution width, that is, as $D^{-1}$.

{\bf Bending and skewing} are most clearly illustrated in terms of the solution 
pressure fields for the points marked on the snaking curve of \reffig{fig:snaking_details}(b). 
The interior structure of the equilibrium solution in \reffig{fig:bendskew}(a) 
is oriented along a diagonal line in the $x,z$ plane, whereas that of the traveling 
wave in \reffig{fig:bendskew}(b) curves upward in $x$ with increasing $z$. We call 
the former effect {\em skewing} and the latter {\em bending}. Skewing is an $x,z$-odd 
phenomenon associated with the $\sxyz$ equilibrium symmetry \refeqn{eqn:twsymm}, 
which gives an odd symmetry $p(x,z) = p(-x,-z)$ in the $y=0$ midplane. Similarly, 
bending is $x,z$-even and associated with the $\tx \sz$ traveling-wave symmetry 
\refeqn{eqn:eqsymm}, which gives an even symmetry $p(x,z) = p(x+\ell_x/2, \, -z)$ 
in the midplane. We quantify skew or bending by the slope ($dx/dz$) or curvature 
($d^2x/dz^2$) of an interpolating function that passes through the local minima and maxima 
of the midplane pressure field. Measured this way, bending and skewing are very nearly 
constant throughout the interior of any given solution, as illustrated by the 
lines of constant slope or curvature in \reffig{fig:bendskew}.

It is notable that the fronts of equilibrium and traveling-wave solutions are indistinguishable at
maximum skew/bend (for example, the right-hand sides near $z\approx20$ in 
\reffig{fig:bendskew}a,b) and also at zero skew/bend (\reffig{fig:bendskew}c,d).
The fronts on the left-hand sides are determined from the right by symmetry. For 
the equilibrium, the odd $p(x,z) = p(-x,-z)$ symmetry means the $dx/dz$ slope of the 
structure has the same sign and magnitude at both the left and right fronts,
so that the two fronts can be connected by a uniform periodic structure with 
constant slope. Importantly, the constant linear slope means the equilibrium solution
can exist two steps higher up in $D$ (width) on the snaking curve, with the
same internal winding structure and the same fronts, simply by adding more 
of the same interior periodic winding structure (or one step by adding half 
as much and flipping the solution with $\sz$). The fact that the equilibrium solution 
can be extended in length this way with no change in interior structure thus 
explains why it snakes in a fixed region of Reynolds numbers, independently of $D$. 

The even $p(x,z) = p(x+\ell_x/2, \, -z)$ symmetry of the traveling wave, on the other 
hand, means that the fronts impose a $dx/dz$ slope with opposite signs at the 
either end, so that the line connecting them generally must curve, as in 
\reffig{fig:bendskew}(b). We observe two features of this curvature in all 
localized traveling-wave solutions. First, the curvature is constant throughout the
solution interior, so that the slope changes uniformly throughout. Thus in 
marked contrast to the equilibrium, the pattern on the interior of traveling wave is not periodic, 
but instead changes smoothly throughout. This is apparent in the changing 
relative streamwise phase of adjacent pressure minima and maxima of the traveling wave in 
\reffig{fig:bendskew}(b), but also more subtly in the long-$z$ modulation of 
the pressure field, which is visible through the changing magnitudes of the 
local maxima and minima of the pressure contours from one end of the solution 
to the other. Second, the slopes at the fronts vary between fixed bounds, 
the same bounds as for equilibria. Consequently, as the solution widens upwards along 
the snaking curve, the curvature decreases, and the interior structure becomes
more periodic.

{\bf Finite-size effects and $D^{-1}$ scaling}. 
\refFig{fig:snaking_threepi}(a,b,c) show bending, skewing, and wave speed as a
function of $D$, in comparison to the $\Rey,D$ snaking in \reffig{fig:snaking_threepi}(d). 
Several features are notable. First, the solutions snake twice as fast in $\Rey$ as in 
skewing, bending, or wave speed. This is due to the fact that the points of maximum 
magnitude in skewing and bending near the upper saddle-nodes in \reffig{fig:snaking_details}(b)
have opposite sign. 
Second, the traveling wave's bending and streamwise wave speed curves are nearly identical 
(\reffig{fig:snaking_threepi}a and c) in all aspects, including position of 
minima, maxima, and zeros, $D^{-1}$ scaling, and remarkably, magnitude. 
Sizable discrepancies between bending and wave speed occur only for $D<40$, when the 
traveling wave consists of only a few copies of the interior periodic pattern 
(e.g. \reffig{fig:velocity}a,b). The nearly identical magnitudes of nondimensionalized 
bending and wave speed holds only for $L_x = 3\pi$; at other $L_x$ the two quantities are 
strongly correlated but differ in magnitude by a factor of two or less.

Third, each of the traveling wave's bending, wave speed, and $\Rey$ snaking plots has a 
$D^{-1}$ envelope, 
whereas the corresponding plots for the equilibrium are constant in $D$. As argued above, the 
constancy of the equilibrium's behaviour in $D$ is due to the fact that, with linear skew, the solution 
can be extended in $z$ and thus bumped up to a higher position on the snaking curve at the 
same Reynolds number simply by adding another copy of the periodic pattern in the interior. 
Thus the equilibrium snakes between constant bounds in Reynolds number and skewing. For the 
traveling wave, on the other hand, if we take the slope $dx/dz$ at the fronts as boundary conditions for constant 
interior curvature $d^2x/dz^2$ over a solution width that scales as $D$, then the curvature 
must scale as $D^{-1} dx/dz$. Given that the slope of the fronts oscillates between fixed 
bounds, the bending then must oscillate between bounds that scale as $D^{-1}$. For 
large $D$ the curvature thus approaches zero, and the interior of the solution approaches 
a constant periodic pattern with skewing, bending, and wave speed approaching zero.

At the point of zero bending (\reffig{fig:bendskew}d), the interior 
structure of the traveling wave is periodic and practically indistinguishable from the structure 
of the equilibrium at zero skew (\reffig{fig:bendskew}c). These points occur near low-$\Rey$ 
saddle-node bifurcations (\reffig{fig:snaking_details}b), suggesting that the reason 
for the close match in the lower bound in $\Rey$ of the equilibrium and traveling-wave 
snaking regions is that the two solutions near the lower bifurcation point differ mainly
in the orientation of one front. Lastly, the complete lack of symmetry in rung solutions 
means that they generally travel in $z$ as well as $x$; however, the nondimensionalized 
$z$ wave speeds are on the order of $10^{-5}$.

\subsection{Core, front, tail structure}
\label{sec:corefronttail}

\begin{figure}
\begin{center}
\footnotesize{(a)} \includegraphics[width=2.4in]{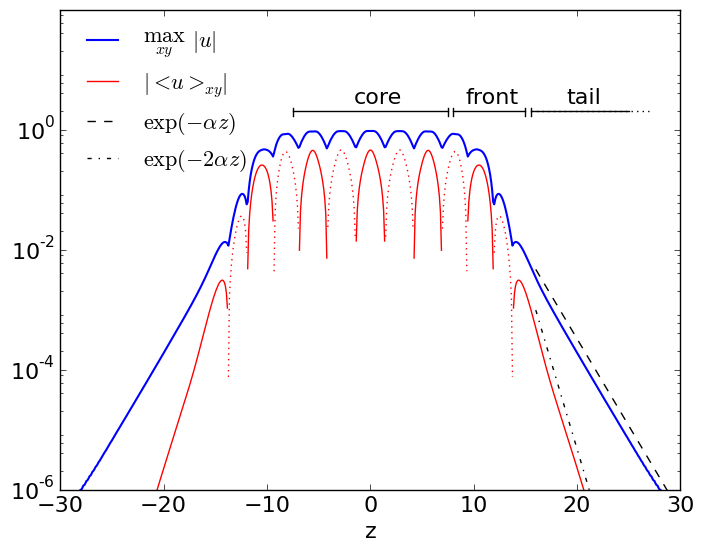}
\footnotesize{(b)} \includegraphics[width=2.4in]{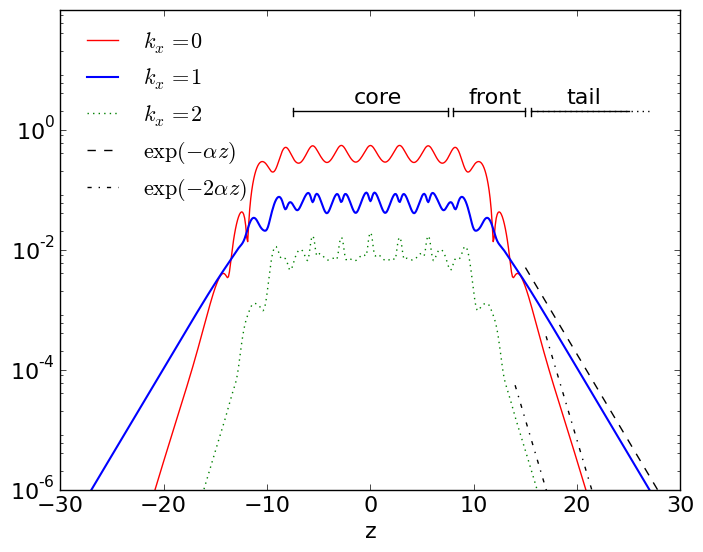}
\caption{
{\bf Core, front, tails structure of the localized solutions}.
(a) Maximum magnitude of streamwise velocity ($\max_{xy} |u|$) and magnitude 
of mean streamwise velocity ($|\xymeanu|$) as a function of $z$ 
for the traveling wave at a point of maximum bend ($\Rey = 275, D=48$). The $|\xymeanu|$
line is dotted when $\xymeanu$ is negative. (b) Magnitude of the $k_x$th streamwise 
Fourier mode for $k_x = 0,1,2$, as measured by the root-mean-square magnitude of 
$\hat{\bu}_{k_x}(y,z)$ over $y$ as a function of $z$. The computational domain 
is $3\pi \times 24\pi$ ($\alpha = 2/3$). 
\label{fig:tails}
}
\end{center}
\end{figure}

The localized solutions are formed from nearly periodic, large-amplitude core
structure surrounded by that taper into small-amplitude, exponentially
decaying tails. The near periodicity of the core and the tapering fronts 
are apparent in \reffig{fig:velocity} and \reffig{fig:bendskew}. \refFig{fig:tails}
illustrates the small-amplitude tails as well, through logarithmic plots of velocity 
magnitude as a function of the spanwise coordinate $z$.  \refFig{fig:tails}(a) shows 
$|\langle u\rangle_{xy}|(z)$, the magnitude of the $xy$-average streamwise flow, 
and $\max_{xy} |u|(z)$, the maximum over $x,y$ of the magnitude of the streamwise 
flow. \refFig{fig:tails}(b) shows the root-mean-square magnitude over
$y$ of several streamwise Fourier modes as a function of $z$. The central feature 
of these plots is the dominant $\exp(-\alpha |z|)$ scaling of the tails (where $\alpha = 2\pi/L_x$), 
consistent with the the linear analysis presented in \cite{GibsonJFM14}. This analysis showed 
that for large $|z|$, the tails of spanwise-localized, streamwise-periodic solutions are 
dominated by the $k_x=\pm1$ streamwise Fourier modes, which take the form 
$\hat{\bu}_{\pm1} (y) \exp{(\pm2 \pi \alpha i (x-c_x t) - \alpha |z|)} + c.c.$. 
The $\exp(-\alpha |z|)$ scaling of the $k_x=1$ mode is apparent in \reffig{fig:tails}(b).
The magnitude of the streamwise velocity in the tails ($\max_{xy} |u|$) is dominated
by the $k_x=\pm1$ modes and thus has the same $\exp(-\alpha |z|)$ scaling, as shown in 
\reffig{fig:tails}(a). The $\exp(-2 \alpha |z|)$ scaling of the $k_x=0$ Fourier mode
in \reffig{fig:tails}(b)
results from a resonance between the $k_x = \pm 1$ modes, which, when summed and 
substituted into the nonlinearity $\bu \cdot \grad \bu$, produce an 
$\exp(-2\alpha |z|)$ forcing term in the $k_x=0$ momentum equation. The $k_x=0$ Fourier 
mode carries the $xy$-average velocity, so $|\xymeanu|$ in \reffig{fig:tails}(a) 
has $\exp(-2 \alpha |z|)$ scaling, as well. The dominant $k_x=1$ mode thus produces 
a small, decaying, but non-zero and constant-sign mean streamwise velocity in the 
solution tails. 

\begin{figure}
\begin{center}
\includegraphics[width=2.4in]{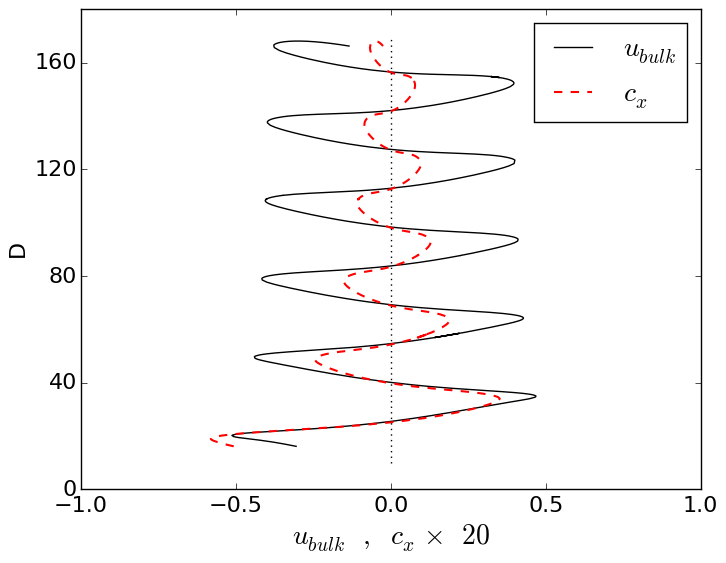}
\caption{{\bf Net streamwise flow and wave speed of the traveling wave.} 
The net streamwise flow $\bar{u} = (L_x L_y)^{-1} \int_{xyz} u \, dx \, dy \, dz$ of 
the traveling wave at $L_x=3\pi$ varies between roughly fixed bounds, in comparison 
to the $D^{-1}$ scaling of the wave speed. Wave speed is magnified by a factor of twenty 
for visibility.
\label{fig:ubulk}
}
\end{center}
\end{figure}

The mean streamwise flow $\xymeanu(z)$ of the traveling wave has a number of interesting 
features due to its $z$-even symmetry, which results from the $\sz \tx$ symmetry of the solution
$\bu$. For one, $\xymeanu(z)$ has the same sign in both tails (positive for the solution 
depicted in \reffig{fig:tails}). Additionally, even $z$ symmetry allows for imbalance between 
positive and negative mean streamwise velocity when summed across the core and front regions.
For example, there are three positive $\xymeanu$ streaks and four negative 
large-magnitude streaks across the core and initial front of the traveling wave in 
\reffig{fig:tails}, flanked by two lower-magnitude positive streaks. Summing across the 
core and fronts, and the weak positive tails, gives a net negative streamwise flow across
the entire computational domain. Thus we have, with zero pressure gradient conditions, a
steady-state solution whose streamwise flow is net negative in the interior, net positive 
in the tails, and net negative over the whole flow domain. \refFig{fig:ubulk} shows how 
the net streamwise flow $\bar{u} = 1/(L_x L_y) \int_{xyz} u \, dx \, dy \, dz$ varies
along the snaking curve (again, the lack of $L_z$ normalization provides for a measure
of the deviation from laminar flow that is insensitive to the computational domain). Note 
that $\bar{u}$ varies between roughly fixed bounds as the solution width ($D$) increases, 
because it results from an $N$ versus $N+1$ imbalance of large-magnitude streamwise 
streaks of opposite sign. This is in contrast to wave speed and bending, which result 
from balances between the fixed-sized fronts and the increasing core and therefore
scale as $D^{-1}$. For the localized equilibrium solution $\bar{u}$ is zero and the streamwise 
flows in the $\pm z$ tails have opposing sign, due to $\sxyz$ symmetry of $\bu$ and consequent 
odd symmetry in $\xymeanu(z)$.

\section{Effects of varying streamwise wavelength}
\label{sec:varyingLx}

\subsection{Snaking region, skewing, and snaking breakdown}
\label{sec:breakdown}





\begin{figure}
\begin{center}
\footnotesize(a) \includegraphics[width=2.4in]{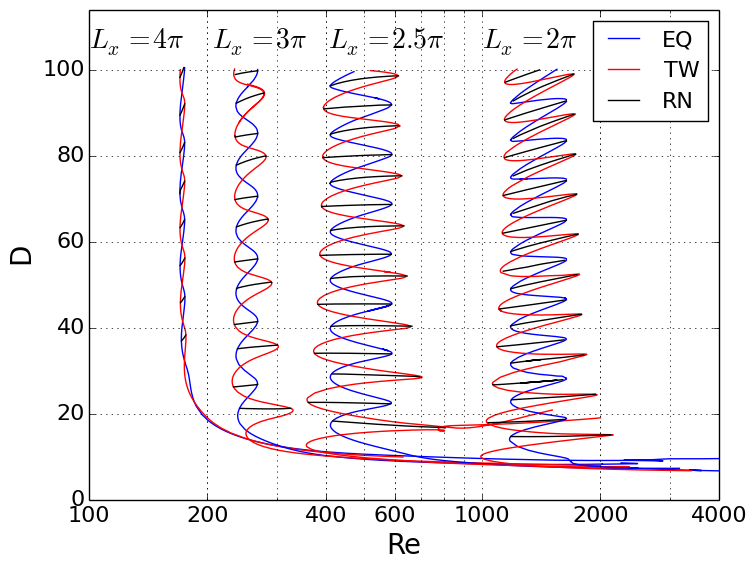}
\footnotesize(b) \includegraphics[width=2.4in]{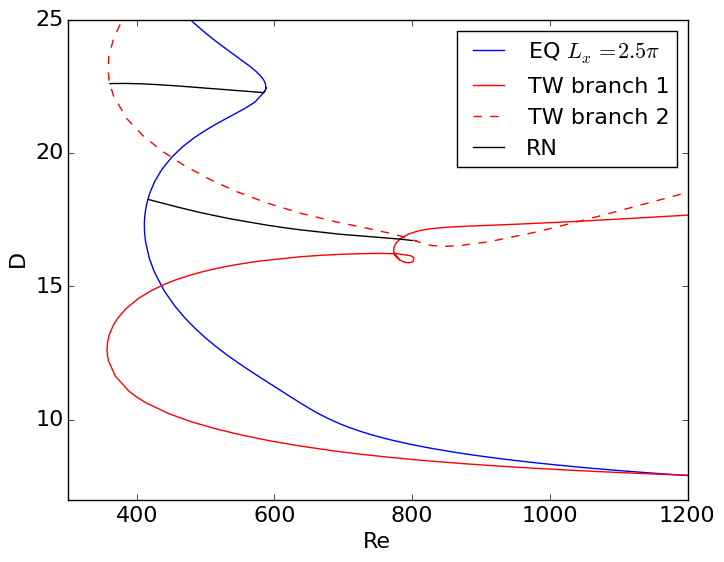} \\
\footnotesize(c) \includegraphics[width=2.4in]{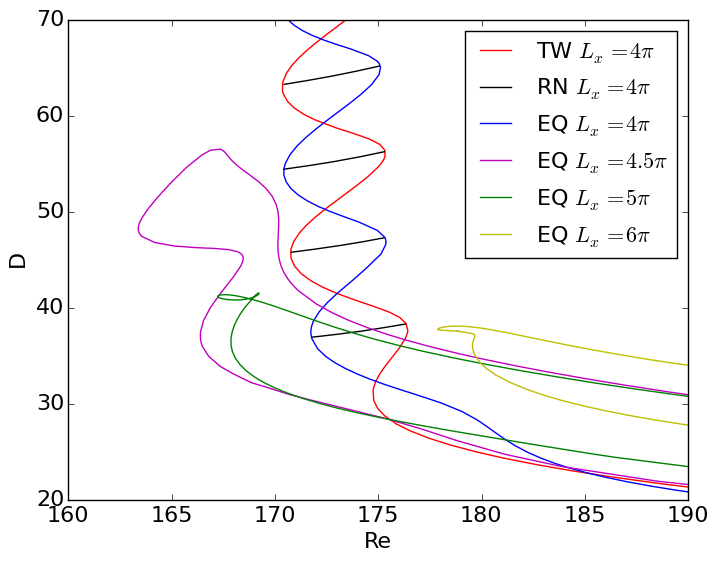}
\footnotesize(d) \includegraphics[width=2.4in]{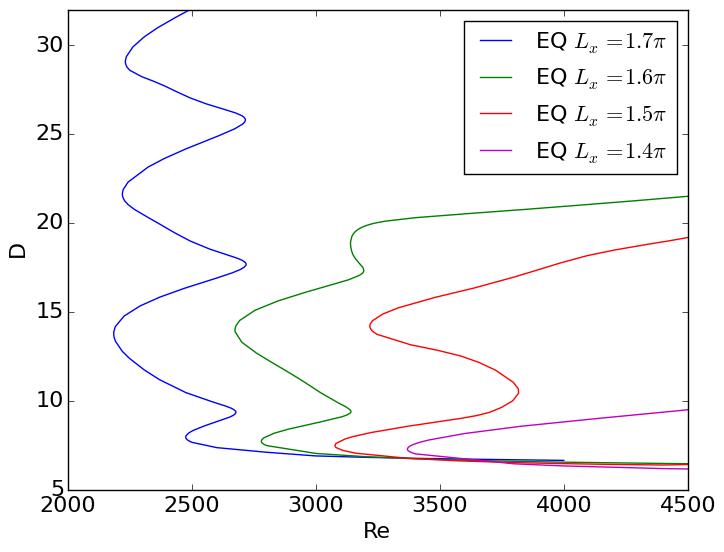}
\caption{{\bf Snaking as a function of streamwise wavelength $L_x$.}
(a) Snaking curves for the equilibrium (EQ), traveling-wave (TW), and rung (RN) solutions at 
$L_x = 2\pi, 2.5 \pi, 3\pi,$ and $4\pi$. 
(b) Detail of snaking curve for $L_x=2.5\pi$. 
(c) Snaking breakdown for $L_x \greater 4.2\pi$. 
(d) Snaking breakdown for $L_x \less 1.7\pi$.
\label{fig:snaking_all}
}
\end{center}

\end{figure}

In this section we examine the effects of changing the streamwise wavelength $L_x$. 
The central results are that snaking is robust in $L_x$ over the range 
$1.7\pi \less L_x \less 4.2\pi$  or $0.48 \less \alpha \less 1.2$, and that 
the snaking region moves upward in Reynolds number with decreasing $L_x$, with snaking 
observed over the range $165 \less \Rey \less 2700$. Each of these bounds is reported
to two digits accuracy. Thus the localized solutions 
and the homoclinic snaking behaviour occur over a wide range of Reynolds numbers, including 
the $\Rey \approx 300$ to $400$ range where \cite{BarkleyPRL05} and \cite{DuguetPF09}
observed laminar-turbulent patterns in plane Couette flow. 
\refFig{fig:snaking_all}(a) shows snaking curves for the localized solutions at a variety 
of streamwise wavelengths. The interlinked snaking structure of the equilibrium, 
traveling-wave, and rung solutions is preserved under variation in $L_x$ with the 
following trends. As $L_x$ decreases, the snaking region moves upwards in $\Rey$ and 
widens. As in the $L_x=3\pi$ case, the width in $\Rey$ of any given equilibrium snaking 
curve is constant in $D$, whereas the amplitude of the traveling-wave snaking region 
decays as $D^{-1}$. For $L_x = 4\pi$ the excess amplitude of traveling-wave snaking 
region over the equilibrium is too small to be observed.


\begin{figure}
\begin{center}
\footnotesize(a) \hspace{-2mm} \includegraphics[width=2.4in]{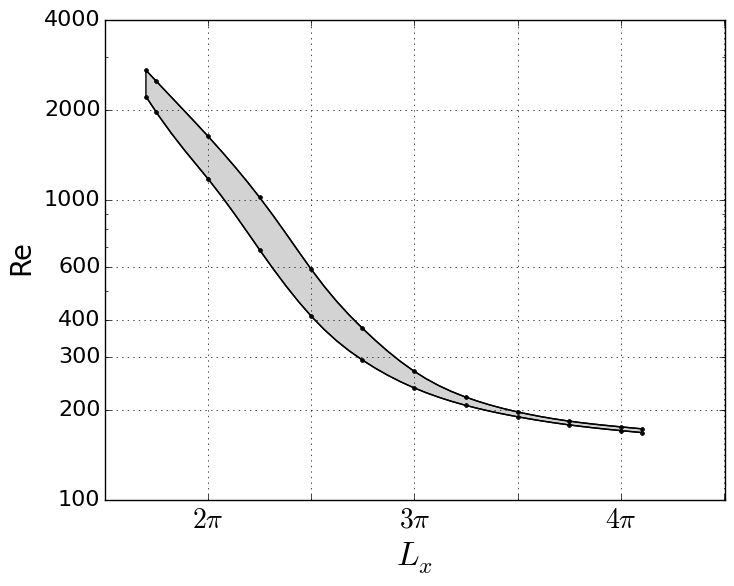} 
\footnotesize(b) \hspace{-2mm} \includegraphics[width=2.4in]{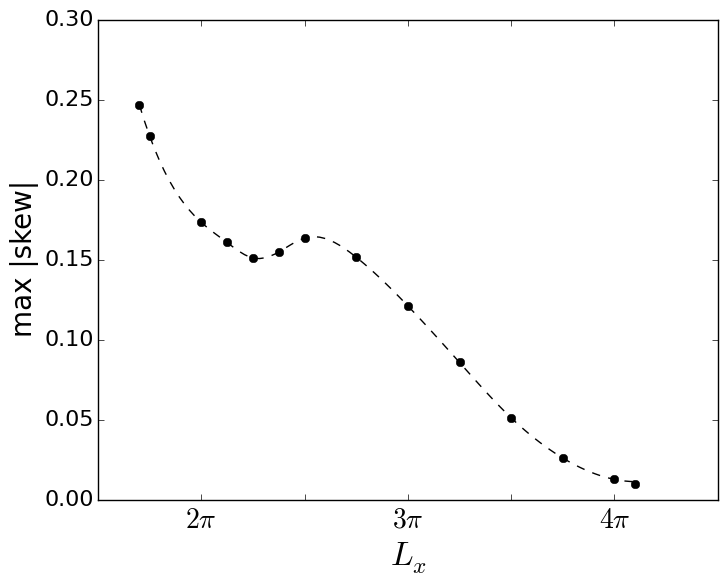} 
\end{center}
\caption{{\bf Snaking region and maximum skewing as a function of wavelength.}
(a) The snaking region in $\Rey$ as a function of $L_x$ for the localized equilibrium. 
The shaded region indicates the range of $\Rey$ within which snaking occurs at a
given $L_x$. 
(b) Maximum magnitude of skewing of the equilibrium solution as a function 
of $L_x$. 
For both figures, dots mark measured values, and the curves are interpolated. 
}
\label{fig:wavelength}
\end{figure}

\refFig{fig:wavelength}(a) shows the snaking region of the localized equilibrium in 
Reynolds number as a function of the streamwise wavelength $L_x$. The boundaries of 
the snaking region in $\Rey$ decrease roughly exponentially with $L_x$ for 
$L_x \less 3\pi$. As $L_x$ increases to $4.2\pi$, the lower bound of the snaking
region approaches a constant $\Rey = 165$. 
Snaking breaks down outside the range $1.7\pi \less L_x \less 4.2\pi$. 
\refFig{fig:snaking_all}(c,d) illustrates the manner of the snaking breakdown.
Above and below this range, the solutions are unable to grow additional structure 
indefinitely at the fronts. Instead of snaking indefinitely, the solution curve 
turns around and continues to higher Reynolds numbers at roughly constant 
solution width. For $L_x$ values just beyond the given range, the solutions snake
a few times before turning around, as illustrated by the $L_x = 1.6\pi$ and $4.5\pi$ 
curves in \refFig{fig:snaking_all}(c,d). 

It is notable that the breakdown of snaking at $L_x \approx 4\pi$ closely coincides 
with the vanishing of the amplitude of the $D^{-1}$ scaling in the traveling wave's 
snaking region, as seen \reffig{fig:snaking_all}(a). 
Similarly, the magnitudes of bending and skewing decrease with increasing $L_x$, 
and at $L_x \approx 4\pi$ are too small to be observed in plots of the velocity or pressure 
fields. \refFig{fig:wavelength}(b) shows the magnitude of the oscillation in skewing over the 
snaking curve as a function of $L_x$, as measured by the slopes of the lines through pressure
minima and maxima, as shown in \reffig{fig:bendskew}(b). 
It is possible that the breakdown of snaking for $L_x \gtrsim 4\pi$ is related to the 
disappearance of these effects at $L_x \approx 4\pi$.

Another defect in the homoclinic snaking scenario is shown in \reffig{fig:snaking_all}(b). 
This detail of the snaking curves for $L_x = 2.5\pi$ shows that the traveling wave solution has 
two distinct branches. The solid curve labeled ``TW branch 1'' was continued downward from high 
Reynolds numbers and small widths ($D \approx 7$) where it connects smoothly to the other snaking 
curves via continuation in $L_x$. However this solution branch does not snake; rather it 
turns around in a saddle-node bifurcation and continues back to at least $\Rey =2000$ at finite width.
In contrast the dashed curve labeled ``TW branch 2'' does snake upward in $D$; it constitutes the 
bulk of the $L_x=2.5\pi$ traveling wave snaking curve shown in \reffig{fig:snaking_all}(a). The 
branch 2 traveling wave was obtained from the endpoint of the rung solution at $\Rey \approx 360$,
$ D \approx 22$. We have observed similar defects in snaking curves at several other values of 
$L_x$ (not shown). It is possible that the snaking breakdown observed for $L_x < 1.7\pi$ and 
$L_x > 4.2\pi$ is of this type. That is, there might be branches of the solution curves
for such $L_x$ that do snake but are disconnected from the solution curves pictured in 
\reffig{fig:snaking_all}(c,d). Lastly, we note that snaking occurs when the solutions are continued 
in $L_x$ with $\Rey$ fixed, within the shaded $\Rey, L_x$ parameter regions depicted in 
\refFig{fig:wavelength}(a).

\subsection{Stability}

\begin{figure}
\begin{center}
\footnotesize{(a)} \hspace{-2mm} \includegraphics[width=2.4in]{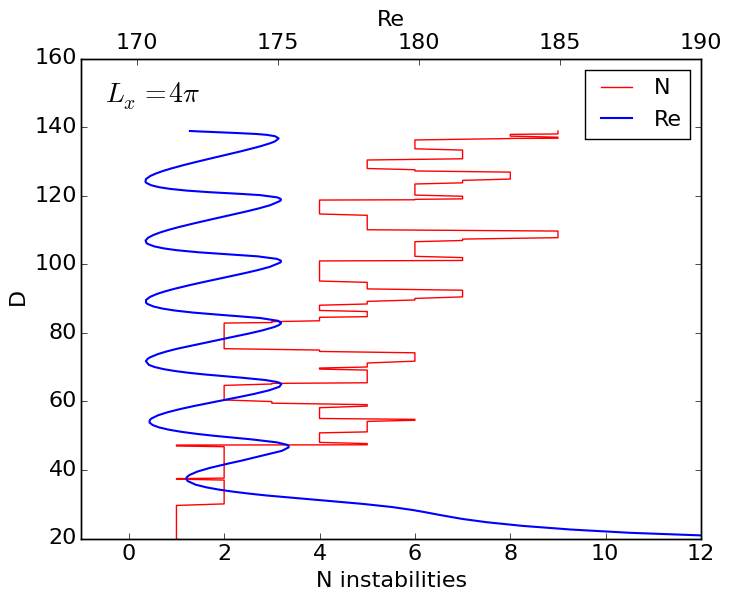} ~~
\footnotesize{(b)} \hspace{-2mm} \includegraphics[width=2.4in]{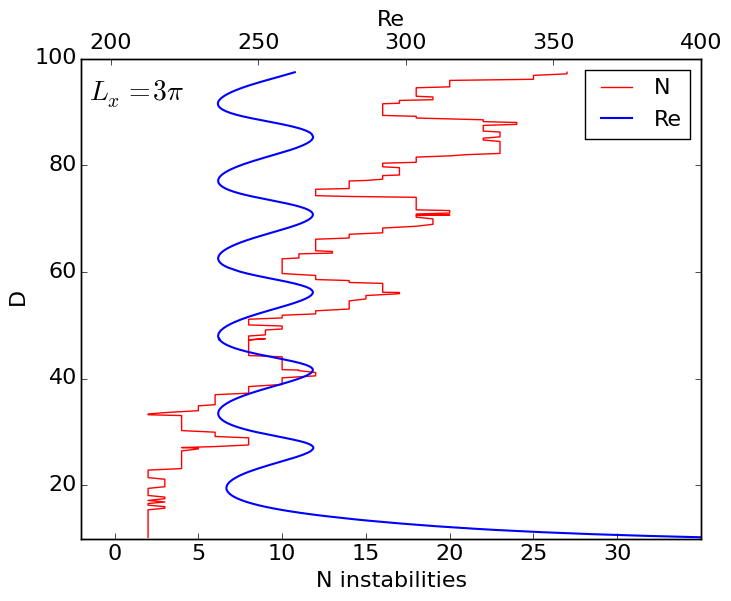}
\end{center}
\caption{{\bf Instability of the localized equilibrium in relation to width 
and wavelength.} The number of unstable eigenvalues of the localized equilibrium 
solution as a function of solution width ($D$), overlaid on the $\Rey,D$ snaking 
curve for (a) $L_x = 4\pi$, (b) $L_x = 3\pi$. 
}
\label{fig:instabilities}
\end{figure}

\refFig{fig:ubulk} shows the number of unstable eigenvalues of the $L_x= 4\pi$ 
and $3\pi$ equilibrium solutions in comparison to their $\Rey, D$ snaking curves. 
At $L_x=3\pi$ the equilibrium has a minimum of two or three unstable eigenvalues 
at small spanwise width (low $D$). Thus it is not strictly an edge state of the 
flow \citep{SkufcaPRL06,SchneiderCH06}. However one of the corresponding unstable eigenfunctions 
is antisymmetric, making the solution an edge state of the flow when constrained 
to $\sxyz$ symmetry. In both cases there is a general trend toward more unstable 
modes as the solution grows wider. The $L_x = 3\pi$, $D \approx 100$ solutions 
depicted in \reffig{fig:bendskew}(a,c) have $O(20)$ unstable eigenvalues. 
Superimposed on this general trend is an oscillation in which the
number of unstable eigenvalues increases and decreases along the snaking curve. 
For both cases the local maxima (minima) in the 
number of unstable modes occur at points of maximum (minimum) skewing
magnitude. In other words, strongly skewed solutions are more unstable 
than solutions with weak or zero skew. The same trends occur at $L_x=2\pi$, with the 
smallest-width solution starting with six unstable eigenvalues. The trend towards more 
instabilities with increasing solution size contrasts with the Swift-Hohenberg 
equation, which shows no such growth. There is an oscillation in the instabilities 
of the one-dimensional cubic-quintic Swift-Hohenberg with each cycle along the 
snaking curve, between one and zero unstable eigenvalues \citep{BurkeChaos07}.

\section{The periodic pattern of the core}
\label{sec:periodicpattern}

\subsection{Relation of the periodic pattern to the NBCW solution}
\label{sec:nagata}

\begin{figure}
\begin{center}
\footnotesize{(a)} \includegraphics[height=1.75in]{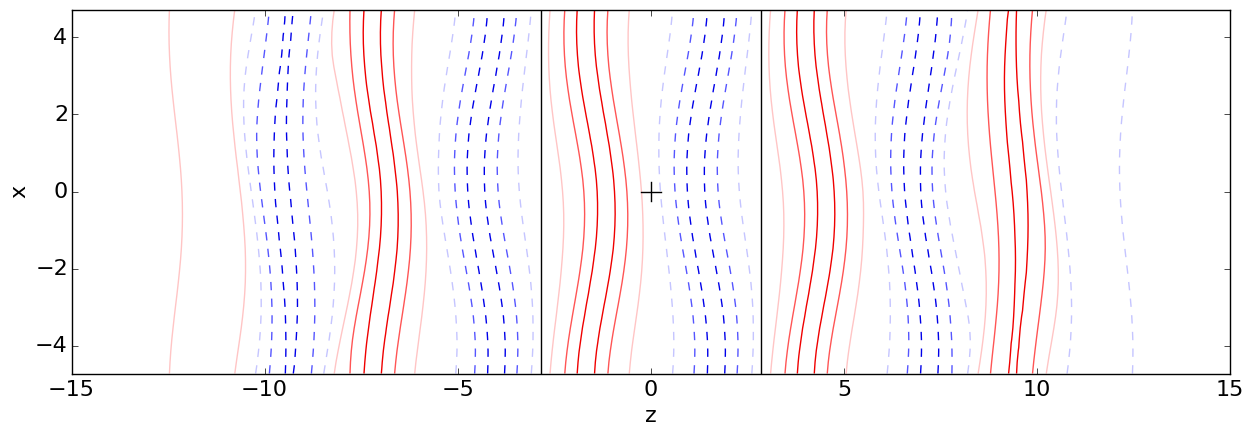} \\
\footnotesize{(b)} \hspace{-2mm} \includegraphics[height=1.75in]{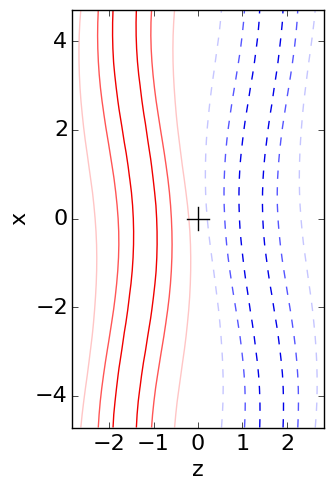}
\footnotesize{(c)} \hspace{-2mm} \includegraphics[height=1.75in]{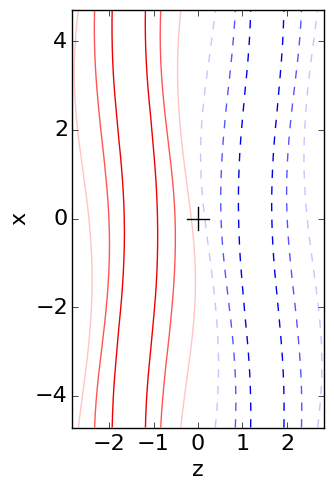}
\footnotesize{(d)} \hspace{-2mm} \includegraphics[height=1.75in]{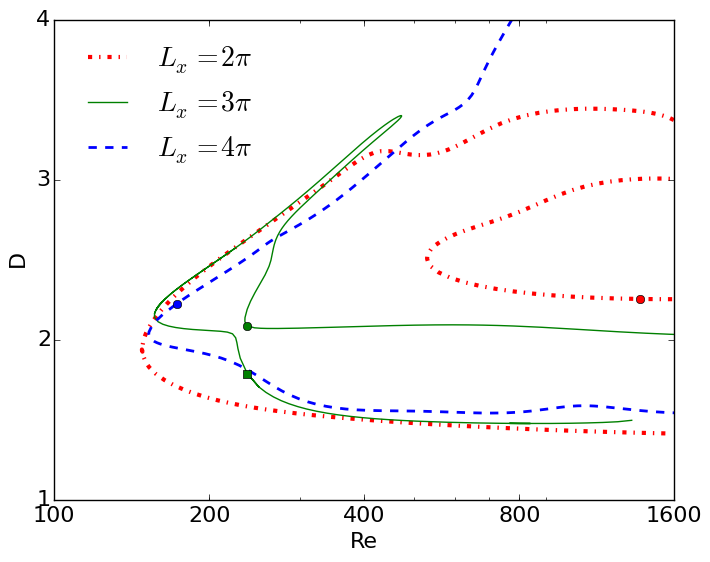}
\caption{
{\bf Extraction of the interior periodic pattern of a spanwise localized solution.}
(a) The spanwise-localized equilibrium at point of zero skew, $L_x = 3\pi$, $D=47$, 
and $\Rey=237$.
Vertical lines at $z = \pm 2.85$ mark one copy of the nearly periodic interior 
structure. 
(b) An exact periodic equilibrium obtained by Newton-Krylov refinement of
the structure extracted from (a), with $L_x, L_z = 3\pi, 5.7$ and  $\Rey=237$.
(c) The lower-branch NBCW equilibrium at the same parameter values as (b), 
obtained by continuation. Plotting conventions for (a,b,c) are the same as in
\reffig{fig:velocity}. 
(d) Bifurcation curves for the interior periodic pattern and the NBCW 
equilibrium for $L_x=2\pi, 3\pi$ and $4\pi$ and aspect ratios 
$L_x/L_z = 1.74, 1.65, 1.74$ respectively. The periodic interior pattern 
and lower-branch NBCW equilibrium shown in (b,c) are marked on the $L_x = 3\pi$ 
curve with a circle and square respectively. The interior pattern extracted 
from spanwise-localized equilibrium at $L_x=2\pi$ and $4\pi$ are also marked 
with circles.
\label{fig:extraction}
}
\end{center}
\end{figure}

\cite{SchneiderJFM10} and \cite{SchneiderPRL10} established that the snaking solutions 
are closely related to the NBCW equilibria, in that they result from a localizing
bifurcation of NBCW and resemble the structure of NBCW in their interior.
However, the relationship between the core structure and the NBCW solutions is more 
complicated than previously supposed. In particular, for $L_x \less 3\pi$, the interior
pattern and the NBCW solution lie on distinct solution curves when continued in Reynolds 
number, although these curves can be connected by continuation in the higher-dimensional
parameter space $L_x, L_z, \Rey$. 

To compare the interior pattern to the NBCW solution, we extracted one copy of the 
interior pattern of the equilibrium at a variety of $L_x$ values, as illustrated 
for $L_x = 3\pi$ in \reffig{fig:extraction}. We begin in \reffig{fig:extraction}(a) 
with a streamwise-localized equilibrium solution at a point along the snaking curve 
of zero skew, in order to maximize the spanwise periodicity of the interior pattern. 
The natural spanwise wavelength $\hat{L}_z = 5.70$ of the interior pattern was determined 
by finding the zeros of $\langle u\rangle_{xy}(z)$ on either side of $z=0$. In 
\reffig{fig:extraction}(a) these points are marked with vertical lines at $z= \pm 2.85$. 
The nearly periodic interior pattern was then interpolated onto uniformly spaced grid 
points for a spanwise periodic computational domain of width $L_z = 5.70$ and refined 
with a Newton-Krylov search to the exact equilibrium shown in \reffig{fig:extraction}(b). 
We performed this operation to find the exact equilibrium solution corresponding to the 
interior periodic pattern at several streamwise wavelengths in the range $2\pi \leq L_x \leq 4\pi$. 
In each case the divergence and the Gibbs phenomenon of the interpolated field were small 
and the Newton-Krylov refinement converged quickly onto an exact equilibrium. The natural
aspect ratio of the interior pattern was always found to be in the range 
$1.65 \leq L_x/\hat{L}_z \leq 1.75$. 

\refFig{fig:extraction}(d) shows bifurcation diagrams for the the NBCW solution 
and exact equilibrium computed from the interior pattern for $L_x = 2\pi, 3\pi$, 
and $4\pi$. Since these solutions are spanwise as well as streamwise periodic,
we use the conventional measure of energy dissipation and wall shear rate
\begin{align}
D_{tot} = I_{tot} = \frac{1}{2 L_x L_z} \int_{-L_x/2}^{L_x/2} \int_{-L_z/2}^{L_z/2} \left. \dd{u_{tot}}{y} \right|_{y=-1} + \left. \dd{u_{tot}}{y} \right|_{y=1} dx \, dz.
\label{eqn:Dtot}
\end{align}
The circles mark the $\Rey, D_{tot}$ positions of the 
interior-pattern equilibrium solutions computed from localized solutions as described 
above, and the lines indicate the parametric continuation of these solutions in $\Rey$. 
For $L_x=4\pi$ it was straightforward to continue the NBCW solution to the same 
aspect ratio and Reynolds number and confirm that the pattern and the NBCW solution 
were the same. However, as $L_x$ decreases, the upper portion of the solution curve 
pinches off at a codimension-2 bifurcation point near $L_x=3\pi$, $\Rey=237$, leaving 
the interior pattern and NBCW on distinct solution curves. This pinching occurs at 
an $L_x$ value just below the $L_x=3\pi$ solution curve shown in \reffig{fig:extraction}(d).
For $L_x=2\pi$, the solution curves for the interior pattern (dash-dot line
marked with a circle) and the NBCW solution (dash-dot, no marker) are distinct. 

\subsection{A winding solution of plane Couette flow}
\label{sec:winding}

\begin{figure}
\begin{center}
\includegraphics[width=5.1in]{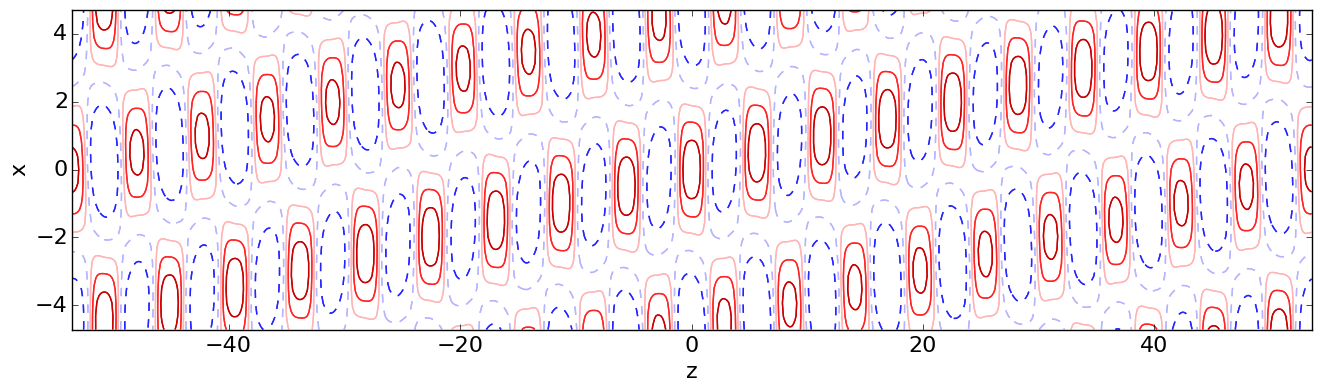}
\caption{{\bf A winding equilibrium solution of plane Couette flow} at $\Rey = 268$.
The solution is strictly periodic in $x$ but has $z$ periodicity involving 
a phase shift in $x$ of the form $\bu(x,y,z+5.642) = \bu(x-0.496,y,z)$. 
Nineteen copies of the $3\pi \times 5.642$ pattern fit in the $3\pi \times 107.2$ 
periodic computational domain. Contours of the midplane pressure field $p(x,z)$ 
are shown with the same plotting conventions as \reffig{fig:bendskew}.
\label{fig:winding}
}
\end{center}
\end{figure}

It is possible to compute a spatially periodic {\em winding} solution from a 
skewed localized equilibrium, that is, a solution with fundamental 
domain size $L_x,\hat{L}_z$ that is strictly periodic in $x$, but whose $z$ periodicity involves a phase shift in $x$, 
\begin{eqnarray}
\bu(x,y,z+\hat{L}_z) = \bu(x-\Delta x,y,z).
\label{eqn:windingsymm}
\end{eqnarray}
One such solution is illustrated in \reffig{fig:winding}. The solution was computed 
starting with a localized equilibrium with strong skewing, like that shown in 
\reffig{fig:bendskew}(a). An iterative process of continuation in the computational 
domain length $L_z$ and adjustment of skewing by continuation in Reynolds number 
was performed to find a localized equilibrium whose interior pattern divided the 
computational domain evenly ($L_z/\hat{L}_z \doteq N$) and whose skew precisely 
aligned with the $x,z$ diagonal of the computational domain. 
$N$ copies of this winding interior pattern were then 
interpolated onto the computational domain as shown in \reffig{fig:winding} and 
refined with Newton-Krylov search. The resulting winding solution shown in 
\reffig{fig:winding} at $\Rey = 268$ has a $L_x = 3\pi, L_z = 107.2$ computational 
domain and $N=19$, giving $\hat{L}_z = L_z/N \doteq 5.462$, $\Delta x = L_x/N \doteq 0.496$, 
and winding symmetry $\bu(x,y,z+5.642) = \bu(x-0.496,y,z)$. The winding symmetry and 
$\bu = \sxyz \bu$ were enforced during the Newton-Krylov search. It is likely that
other winding solutions of shear flows could be computed without recourse to 
skewed localized solutions, simply by applying skewing coordinate transformations to 
known spatially periodic solutions and refining with Newton-Krylov search.

\section{Conclusions}
\label{sec:conclusion}

We have shown that homoclinic snaking is robust under changes in streamwise 
wavelength for the spanwise-localized solutions of plane Couette flow of 
\cite{SchneiderPRL10}. Homoclinic snaking occurs for these solutions over 
streamwise wavelengths in the range $1.7\pi \less L_x \less 4.2\pi$ and 
Reynolds numbers $165 \less \Rey \less 2700$, and the snaking region 
moves upwards in $\Rey$ as $L_x$ decreases. The localized equilibrium, traveling-wave, 
and rung solutions thus exist at arbitrarily large spanwise widths over a wide range 
of Reynolds numbers. Several new properties of the solutions 
become apparent as $L_x$ decreases below $4\pi$, most importantly the linear 
skewing of the equilibrium and quadratic bending of the traveling wave. The traveling 
wave exhibits finite-size effects such as $D^{-1}$ scaling of the bending, wavespeed, 
and snaking region, due to the nonuniform structure in the solution core induced by 
the quadratic bending. The linear skewing of the localized equilibrium solution,
on the other hand, induces no such finite-size effects. Its core region is very 
nearly periodic, close enough that a strictly periodic winding solution can be
easily developed from it. The number of instabilities of the localized solutions 
increase with Reynolds number, with width, and with skewing. Thus, at a fixed Reynolds 
number, from a statistical viewpoint one would expect narrow patches of the localized
pattern to appear more frequently than wide patches, and with weak rather than 
strong skewing. 

The homoclinic snaking of these localized solutions suggests the Navier-Stokes 
equations might be related to the Swift-Hohenberg equation under plane Couette 
flow conditions and for certain parameter ranges and flow states. A primary 
motivation for this paper is to clarify the parameter ranges and solution 
structures for which this connection might occur. Our results indicate that homoclinic 
snaking is a finite-Reynolds, finite-wavelength effect, and that the streamwise 
wavelength and the Reynolds-number snaking region are strongly coupled. Thus it 
is unlikely that an analytic understanding of homoclinic snaking in shear flows 
will be found via asymptotic analysis in large-Reynolds or large-wavelength limits. 
If the spanwise-localized solutions are to be understood as a long-wavelength
modulation of a small-wavelength, spanwise-periodic pattern, our results 
show that the periodic pattern is a form of the NBCW solution at aspect ratio 
$L_x/L_z \approx 1.7$. For $L_x < 3\pi$, under continuation in Reynolds number, 
the periodic pattern lies on a solution curve distinct from the widely-studied 
NBCW lower-branch solution.


We see no clear connection between the skewing of the localized equilibrium 
solution and the skewed laminar-turbulent patterns observed in plane Couette 
flow by \cite{BarkleyPRL05} and \cite{DuguetPF09}. The localized solutions 
exists over a much wider range of Reynolds numbers than the $300 \less \Rey \less 400$ 
range of the observed laminar-turbulent patterns. 
Also, the skewing of the localized equilibrium is a streamwise phase shift of 
the interior pattern as function of spanwise coordinate, whereas the 
skewing of observed laminar-turbulent patterns
is in the orientation of the 
boundary between turbulent patches and surrounding laminar flow. 
The localized solutions studied here are strictly streamwise periodic, 
and thus have laminar-turbulent boundaries aligned with the streamwise direction. 
One potential route to finding invariant solutions with skewed laminar-nonlaminar
boundaries would be to find a streamwise-localized form of a winding 
solution like that described \refsec{sec:winding}. However, the skewing 
observed in laminar-turbulent patterns is much larger than any observed here. 
\cite{DuguetJFM10} reports that the boundaries of turbulent patches in 
$\Rey \approx 300$ plane Couette flow lie within the range of angles
$\beta = 36^\circ \pm 10^\circ$ off the streamwise axis. This corresponds 
to range of skewing slopes $dx/dz = \tan(90^\circ - \beta)$ between 
roughly 1 and 2, much larger than the maximum skew of $dx/dz \approx 0.1$ 
of the equilibria in the parameter range $\Rey \approx 300$, 
$2.75\pi < L_x < 3\pi$. 

{\bf Acknowledgements:} The authors thank John Burke and Edgar Knobloch for 
illuminating conversations, and John Burke for assistance in the numerical 
calculations. TMS was supported by the Swiss National Science Foundation 
under grant number 200021-160088. Computations were performed on Trillian, 
a Cray XE6m-200 supercomputer at the University of New Hampshire supported 
by the Major Research Instrumentation program of the United States National 
Science Foundation under grant PHY-1229408.



\bibliographystyle{jfm}
\bibliography{snaking}

\end{document}